\documentclass[preprint2]{aastex}

\shorttitle{Position Stability}
\shortauthors{Fomalont et al.}

\begin{document}


\title{The Position/Structure Stability of Four ICRF2 Sources}


\author{Ed Fomalont}
\affil{National Radio Astronomy Observatory, Charlottesville, VA 22903 USA}
\email{efomalon@nrao.edu}

\author{Kenneth Johnston, Alan Fey and Dave Boboltz}
\affil{United States Naval Observatory, Washington, DC 20392, USA}
\email{kenneth.j.johnston@navy.mil; afey@usno.navy.mil; dboboltz@usno.navy.mil}

\and

\author{Tamoaki Oyama and Mareki Honma}
\affil{National Astronomical Observatory of Japan, Mitaka, Tokyo 181-8588, Japan}
\email{t.oyama@nao.ac.jp, mareki.honma@nao.ac.jp} 



\begin{abstract}
Four close radio sources in the International Celestial Reference
Frame (ICRF) catalog were observed using phase referencing with the
VLBA at 43, 23 and 8.6 GHz, and with VERA at 23 GHz over a one year
period.  The goal was to determine the stability of the radio cores,
and to assess structure effects associated with positions in the ICRF.
Although the four sources were compact at 8.6 GHz, the VLBA images at
43 GHz with 0.3-mas resolution showed that all were composed of
several components.  A component in each source was identified as the
radio core using some or all of the following emission properties:
compactness, spectral index, location at the end of the extended
emission region, and stationary in the sky.  Over the observing
period, the relative positions between the four radio cores were
constant to 0.02 mas, the phase referencing positional accuracy
obtained at 23 and 43 GHz among the sources, suggesting that once a
radio core is identified, it remains stationary in the sky to this
accuracy.  Other radio components in two of the four sources had
detectable motion in the radio jet direction.  Comparison of the 23
and 43 GHz VLBA images with the VLBA 8.6 GHz images and the ICRF
positions suggests that some ICRF positions are dominated by a moving
jet component; hence, they can be displaced up to 0.5 mas from the
radio core, and may also reflect the motion of the jet component.
Future astrometric efforts to determine a more accurate quasar
reference frame at 23 and 43 GHz and from the VLBI2010 project are
discussed, and supporting VLBA or EVN observations of ICRF sources at
43 GHz are recommended in order to determine the internal structure of
the sources.  A future collaboration between the radio (ICRF) and the
optical frame of Gaia is discussed.
\end{abstract}





\section{Motivation for the Experiment}
\label{sec:motivation}

The Second Realization of the International Celestial Reference Frame
(ICRF2) is defined by the positions of 295 compact radio sources
located around the sky \citep{icrf2}.  The positional errors for many
of these sources is less than 0.06 mas obtained by averaging decades of
observations, and they define the axes of a radio-based quasi-inertial
frame to about 0.01 mas precision.  This positional accuracy is
presently limited by a typical residual variable tropospheric
refraction of about 20 psec (~0.15 mas) for a time-scale of tens of
minutes and angular size scale of tens of degrees.

The radio source emission for most sources in the ICRF2 catalog
(quasars and galaxies) is dominated by a strong compact radio
component (called the radio core) with a size $<0.1$ mas.  This radio
core is probably located at the base of a radio jet that is formed
perpendicular to the accretion disk about 1000 to 10000 Schwarzschild
radii from the massive object in the galactic nucleus \citep{mar06}.
For quasars and radio galaxies at cosmological distances, this offset
is less than about 0.01 mas, so the radio core should represent a
fixed point in the sky at this precision.  However, there is often
more extended emission (generically called the jet) that may be
connected to the core, or consist of discrete components.  The
emission of the core and jet both vary with time, and the jet often
contains components that move away from the core.  Because the
emission near the core becomes more opaque at lower frequencies, the
determination of the core position becomes problematic.  Although the
magnitude of this uncertainty caused by structure is less than that
from troposphere refraction, it is becoming an important part of the
position error budget as the tropospheric modeling and measurements
improve.

In order to determine the effect of structure and its changes on the
apparent radio core positions obtained from radio observations, the
effect of the tropospheric refraction variations must be significantly
decreased in order to reach the structure astrometric effects.  By
using a VLBI technique called phase referencing, where several sources
that are closely spaced on the sky are observed simultaneously or in quick
succession, a large part of the tropospheric refraction can be
canceled.  For frequencies as high as 43 GHz, the needed phase
stability (residual tropospheric-induced and other phase errors should
be less than about 1 radian) can be obtained most of the time if the
separations among the sources are less than a few degrees, and the
switching times between sources is less than about one minute.

\section{Observations and Initial Data Processing}
\label{sec:observations}

\subsection {Source Selection}

\begin{figure}[hbt!]
\includegraphics[angle=0,scale=0.5]{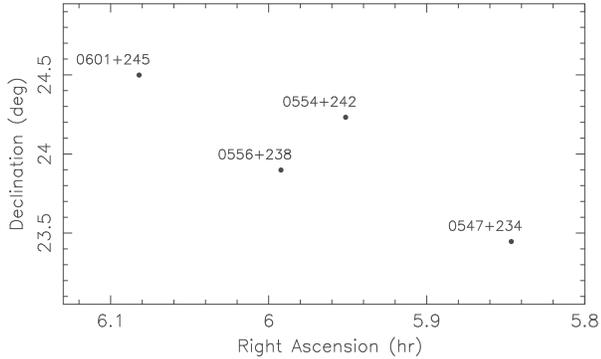}
\caption{The location of the four ICRF sources used for these observations}
\label{fig:configuration}
\end{figure}

The source selection was made to accommodate observations by the VLBI
Exploration of Radio Astrometry (VERA) four-element VLBI array in Japan
\citep{VERA} that can observe two sources simultaneously if they are
separated by no more than $2.2^\circ$.  A search of the 8.6-GHz ICRF
catalog found several sets of sources and the best set for phase
referencing were the four sources: 0547+234, 0554+242, 0556+238 and
0601+245.  The relative position of the four sources are illustrated
in Fig.~\ref{fig:configuration}.  The positions and approximate
intensity parameters for the sources are shown in
Table~\ref{tab:source_parameters}.  The apriori positions, taken from
the ICRF catalog, are fiducial locations for each source (phase center
used in the correlation of the data) to which the measured offset
positions and structures are referred.  The approximate correlated
flux densities at each frequency are given for the projected spacing
of 200 km (S) and at 5000 km (L).  The values have been taken from
these observations.  All of the sources are somewhat variable.
Because the sources are within $5^\circ$ of the galactic plane, there
are no firm identifications, and their redshifts are not known.  The
sources are probably quasars with a redshift between 0.5 to 2.0
\citep{har90}.

The source 0556+238 is one of the 295 {\it defining} sources in the
ICRF2 catalog.  The other three sources are in the extended ICRF2
list, but are not one of the defining sources because of their
somewhat limited use in the ICRF sessions over the past thirty years.

Although this quartet of sources was chosen because of their proximity
in the sky and their relative strength and compact structure at 8.6
GHz, other deeper properties of these sources---internal structure and
its frequency dependence, and variability time-scale---were unknown at
the higher frequencies and resolutions used for these observations.
These four sources are a representative, although limited, sample of
good quality ICRF sources for the deeper investigations reported in
this paper.

\subsection {Observing Sequences}\label{sec:observing_sequence}

The list of observing sessions are given in Table \ref{tab:sessions}
The initial observing request was submitted to the Japanese VERA
array, and the first observing session at 23 GHz occurred on 18 April
2008.  The session was 6.5 hours in length, and cycled among the
source pairs: 0547-0554; 0547-0556; 0554-0556; 0554-0601; 0556-0601,
all with a separation less than $2.2^\circ$ so VERA could observe each
source-pair simultaneously.  Only the source pair separation 0547+234
to 0601+245 was over this angular limit.  Each pair was observed for
an 8-minute scan, and then each of the five pairs had 8 such
observations over each session.  These data enabled accurate images
and pair separations to be determined.  After the second VERA session
on 24 May 2008, an observing proposal was submitted to the VLBA in
order to observe at somewhat higher resolution and at 43 GHz since the
preliminary VERA results indicated the complex nature of the sources.

Six VLBA sessions (see Table \ref{tab:sessions} at 23 GHz and 43 GHz
were scheduled between 20 Dec 2008 and 12 Dec 2009.  For the most
accurate phase referencing stability, a primary reference source that
is observed every other scan was chosen.  The source 0556+238 was
chosen as the primary calibrator because it was the most compact
(ratio of L/S from Table ~\ref{tab:source_parameters} was largest),
and it was relatively strong.  This permitted the monitoring of fast
phase changes caused by the tropospheric refraction variations.  The
other sources were then alternated observed in turn with the primary
phase calibrator.  This observing/reduction scheme uses the primary
reference source to determine the instrumental/atmospheric/astrometric
errors in its direction and assumes that the same errors apply to the
other sources (targets).  The positional reference frame is also tied
to the apriori position of the primary reference source.

The observing sequence used was
0556--0547--0556--0554--0556--0601--0556--0547...).
Each scan was 18 sec in length, with 10 sec needed for source
switching. The separation between the center of two 0556+238 scans was
only 55 sec, sufficiently close in time to maintain phase coherence at
43 GHz except under adverse weather conditions or at very low
elevations.  The basic sequence took 3 min, and ten sequences were
made over a 30 minute period.  These 30-minute blocks were then
alternated between 23 GHz and 43 GHz observing frequencies.

For 40 minutes near the middle of each VLBA session, 1-min scans at 23
GHz of strong ICRF2 sources over the sky were made in order to
determine the residual tropospheric refraction model in addition to
that used for the baseline correlator model.  This observing
technique is called \lq\lq DELZN\rq\rq, and is described by
\cite{DELZN}.

The final VLBA session on 31 Jan 2010 for one hour at 8.4 GHz, used
the same scan cycle in order to obtain the relative source positions
and structure at this lower frequency at the end of the monitoring
period.

\subsection {Data Correlation and A-priori Processing}\label{sec:data_correlation}

The correlator models used for the VERA and VLBA observations used
accurate locations for the antennas, up-to-date earth orientation
(EOP) parameters, and initial tropospheric delay models.  The
positions listed in Table~\ref{tab:source_parameters} are the assumed
positions (phase center) for the sources for all sessions.  These
positions are merely the reference point from which offset positions
are determined.  They are often close to the location of the maximum
intensity of the radio source.

For all VERA sessions, the correlated data were sampled every 2 sec,
with 512 channels in one frequency band, covering at total bandwidth
of 128 MHz, centered at 23.564 GHz.  After flagging data associated
with mal-functioning antennas or system temperatures in excess of 200K
(the nominal operating system temperature of 100-120K increased
significantly during periods of poor weather), the remaining data were
then averaged to 6-sec intervals and 32 channels covering the 128 MHz
bandwidth.  No bandpass corrections in phase and amplitude were
needed.  Each of the five source pairs for each session were processed
separately: 0547-0556(PR), 0554-0556(PR) 0601-0556(PR), 0547-0554(PR),
0601-0554(PR); where (PR) signifies the source that was used as the
phase reference.  The first three pairs contain all of the information
about the relative source positions among the three sources with
respect to 0556+238, and it is these data that were used in order to
be directly comparable to the VLBA observing strategy.  However, the
observations for the last two pairs are independent, and provide an
estimate of the errors associated with the source positions and
structures.

For all VLBA 23 and 43 GHz sessions, the correlated data were sampled
every 1 second, with a total bandwidth of 64 MHz (2-bit sampling;
except for 20 and 22 Dec 2008 with 1-bit sampling) contained in four
contiguous frequency bands, each with sixteen 1-MHz channels.  The
center frequencies were at 22.45 GHz and 43.41 GHz.  At 23 GHz the
amplitude and phase bandpass over the frequency range were determined
from the observations of 0556+238, and data that were associated with
mal-functioning antennas also were flagged.  At 43 GHz, we inserted
two observations of the strong ICRF2 source, 0851+202, in order to
determine the bandpass amplitude and phase at this frequency.  For the
final VLBA session at 8 GHz of one hour, the same frequency and
correlator configuration were used, and the average frequency was 8.64
GHz.

  Updates of telescope positions, improved EOP, phase compensation for
  the parallactic angle changes over the session, and corrections from
  the monitored antenna electronic gain and phase variations were
  applied to the correlated data.  For the VERA observations, the
  phase and delay difference between the dual-beams were measured
  using an internally-generated broad-band noise source, and these
  corrections were applied.  The offsets are large, although
  relatively stable with time, and are critical corrections needed for
  precise phase referencing.

The estimates of the tropospheric delay for the VERA data were
obtained from GPS measurements at each antenna.  These are accurate to
about 1cm, except at the more humid locations at Ogasawra and
Ishigaki-jima \citep{hon08}.  In order to improve the tropospheric
delay model used in the VLBA correlator, analysis of the DELZN
observations, taken near the mid-point of each session, determined the
zenith-path delay offset for each antenna \citep{DELZN}.  The
frequency configuration consisted of four bands each with 16 1-MHz
channels, centered at 23.02 23.57, 23.29 and 23.46 GHz.  The
antenna-based group delays were measured, and then the data for each
antenna were fit to a linear clock delay and a zenith-path delay
residual over the observation period of 40 minutes \citep{aips-DELZN}.
These results are not as accurate as the GPS-measured tropospheric
delays, but decreased the systematic zenith-path delay model errors
for each VLBA antennas from about 4 cm, to about 1 cm.  At 43 GHz,
zenith-path delay error of 1 cm for a source pair two degrees apart
give a systematic phase error of $18^\circ$ at the zenith and
$30^\circ$ at a source elevation of $50^\circ$.  When this error is
averaged over all antennas, the systematic position shift from this
uncertainty is estimated to be about 0.01 mas (reference to meeting in
Socorro).
 
These data were then ready for phase-referencing calibrations.

\section{Phase Reference Calibrations}

\subsection {The Primary Calibrator 0556+238}\label{sec:0556_choice}

The choice of 0556+238 as the primary reference calibrator is
discussed in \S~\ref{sec:observing_sequence}. Its ICRF history
suggested that it is relatively stable over time, and was included as
a defining sources in the ICRF2 catalog since its position varied by
less than 0.3 mas in both alpha and delta at 8.6 GHz over several
decades.  Although $>80\%$ of the emission is contained in a small,
unresolved radio component, images of the source were made for each
frequency and session in order to determine the small structure phase.
The source peak brightness is assumed to be located at the position
given in Table~\ref{tab:source_parameters} at both frequencies and for
all sessions, and this defines the relative position grid for the
entire experiment.

It will turn out that the assumption---that the location of the peak
intensity for this source is fixed over the year period of this
experiment---is incorrect at the level of nearly 0.2 mas!  This is
significantly less than the image resolution, but the source position
change with time was easily determined from the anomalous motions of
the other sources.

\subsection {VERA Phase Referencing}

\begin{figure}[hbt!]
\includegraphics[angle=0,scale=0.4]{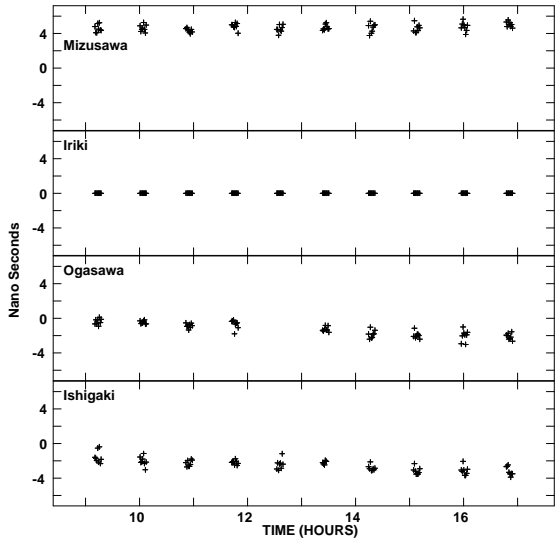}
\caption{The Delay Solutions for the VERA 17-June-2009 session.  For
  each 8-min scan of 0547+234, a delay solution every 1 min was
  obtained.  Mizusawa is in northern Honshu, Iriki is the reference
  antenna and is located 1000 km wast of Tokyo.  Ogaswara is located
  1000 km south of Tokyo, and Ishigaki is located 200 km east of
  Taiwan.}
\label{fig:VERA_delay}
\end{figure}

Since both phase calibrator and target are observed simultaneously
with VERA observations, phase referencing requires the transferal of
the measured antenna-based temporal phase and gain determined for the
calibrator directly to the target.  For each 8-min scan on a source
pair, the AIPS task FRING determined the residual phase, and phase
derivative with time (rate) and frequency (delay) for each one-minute
interval during a scan \citep{AIPS}.  An example of the delay solutions for the
a VERA session is shown in Fig.~\ref{fig:VERA_delay}.  The delay
values were slowly changing with time with an rms scatter over a
seven-minute scan of about 1.0 nsec, caused in roughly equal amounts
by the limited signal-to-noise over one minute and by fast changes in
the tropospheric refraction.  Over the bandwidth of 128 MHz, this
delay scatter corresponds to a phase change of $30^\circ$ across the
band, which causes a decorrelation of only about 4\%.  Delay solutions
that were outside of the range or with significantly increased scatter
were flagged.  These data usually correspond to periods of poor
weather conditions.

After the phase calibrations tables (delay, phase-rate and phase) were
then applied to the phase calibrator, The AIPS task CALIB was then run
on this data, using the input model of 0556+238, to determine the gain
variations for each antenna.  Variations of about 10 to 20\% over the
6.5-hour session were typical.  The phase calibration tables and the
antenna-gain corrections found for the phase calibrator were then
applied to the target source in order to obtain calibrated data.
Since both sources were observed simultaneously, no temporal
interpolation of the corrections found for the calibrators were needed
to apply to the target.

\subsection {VLBA Phase Referencing}

From the \lq\lq DELZN\rq\rq 40-min observations near the mid-point of
each session (\S~\ref{sec:data_correlation}) the residual zenith-path
delay offset and clock delay for each antenna were obtained.  These
corrections were applied to all data (23 and 43 GHz) for the session
to remove the relatively large offset troposphere and clock delay
error.

The VLBA phase referencing analysis, using the observations of
0556+238, were similar to those described above for the VERA session.
These calibrations were done independently for 23 and 43 GHz.  Because
of the excellent stability of the VLBA and the relatively dry
locations of most of the antennas, the AIPS task FRING determined only
the delay and phase for each 20-sec scan of the calibrator.  The
delays were then smoothed over a time scale of one-hour to decrease
the variations from the signal-to-noise.

\begin{figure}[hbt]
\vskip -3cm
\includegraphics[angle=0,scale=0.4]{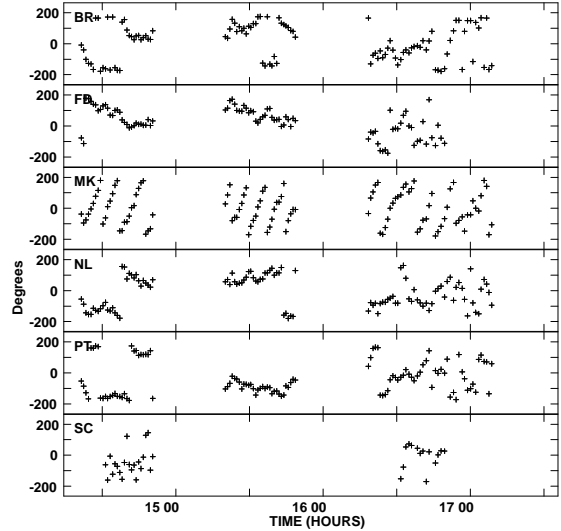}
\caption{The Phase Stability at 43 GHz for observations on 14-Sep-2009
  for the 0556+256 scans between UT 14:20 to 17:20 for selected
  antennas.  Each point is the average antenna-based phase for an
  18-sec scan of 0556+238, with LA as the reference antenna.  Periods
  of poor phase stability between adjacent scans are: nearly all data
  after UT 16:45, SC between 14:30 and 15:00.  The large phase slope
  for MK can be followed between calibrator scans.}
\label{fig:VLBA_phase}
\end{figure}

The antenna-based phase variations for the calibrator over a session
varied considerably, depending primarily on the weather conditions.
Examples of these phase variations for several VLBA antennas at 43 GHz
observations on 08 June 2009 over a three hour period are shown in
Fig.~\ref{fig:VLBA_phase}.  In order to maintain the necessary
temporal phase coherence, the observed antenna-based phase must be
accurately interpolated between adjacent 0556+238 scans, 55 seconds
apart, in order to transfer them to the intervening target scans.  In
Fig.~\ref{fig:VLBA_phase} periods of poor phase stability are
obvious (SC 14:30 to 14:50, FD after 16:20, NL after 16:50), and the
data for the targets during these periods were flagged.  As with the
VERA observations, these phase calibrations were applied to 0556+238
and then the average antenna-based gain calibrations were determined
using the AIPS task CALIB.  The delays, phase and gains from 0556+238
are then interpolated between scans in order to apply them to the
target source.  The target sources were now ready for imaging.

\section{Radio Source Images}

\subsection {Imaging Techniques and Astrometry Analysis}

Images for each source, frequency, and array were then made from the
VERA and VLBA data.  All image sizes were $1024\times 1024$ pixels,
with a pixel separation of 0.01 mas.  The data were weighted by the
inverse-square of the antenna noise, and an additional weighting
(called robust=0) of the data as a function of was data resolution
emphasized the fine-scale structure without significant loss of
signal-to-noise.  The dirty images were cleaned using small boxes
around the areas of known emission.  Because of the substantial
differences in the (u-v) coverage among the individual VLBA sessions
and the VERA sessions due to flagged data or missing antennas, we used
a conservative resolution for the restoring clean beam in order to
produce images with the same resolution.  The following resolutions
(north/south by east/west Full-Width at Half-Power [FWHP] in mas) were
adopted: for the VLBA at 23 GHz, the resolution was $1.0\times 0.5$;
for the VLBA at 43 GHz, the resolutions was $0.6\times 0.3$; for the
VLBA at 8.6 GHz the resolution was $3.0\times 1.5$; and for the VERA
23 GHz, the resolution was $1.0\times 1.0$.  Although many radio
components in these diagrams appear to have an angular size that is
similar to that of the beam, even a small broadening is indicative of
internal structure, and both imaging and visibility model fitting can
determine more precise component parameters and sizes (see
\S~\ref{sec:super_resolution}).

The target VLBA images (0547+234, 0554+242, 0601+245) are affected by
phase noise introduced by the phase-referencing calibration
interpolation errors between calibrator scans, that are implied in
Fig.~\ref{fig:VLBA_phase}.  The VERA observations are not affected by
interpolation errors.  Also, angular decorrelation (the tropospheric
phase refraction between calibrator and target are somewhat different
because of their separation in the sky) corrupts the target phases for
both the VERA and VLBA observations.  Unless these phase errors are
larger than about one radian (all such data are flagged before images
are made), the phase-referenced (PR) images for all sources are on the
reference frame tied to the apriori position of the phase-reference
calibrator to an accuracy that can be a good as 0.02 mas over a few
degrees on the sky \citep{fom05}

The image quality of the target sources can be improved using
self-calibration techniques.  By starting with the PR image as the
estimate of the source structure and position, the self-calibration
algorithm then determines improved antenna-based phases that are
consistent with the PR image.  The improved antenna-based phases are
determined every one minute, and corrected visibility data are
obtained.  This technique is robust because it utilizes data from all
baselines (45 maximum for the VLBA) in order to obtain antenna-based
phase corrections (9 maximum for the VLBA).  See \cite{self-cal} for
more information about self-calibration and its stability.

The self-calibrated (SC) images have better quality than the original
PR images.  The peak brightness of the stronger components increase
from a few percent to 20 percent because the corrected visibility data
have smaller phase errors, and the extended emission and fainter
components are somewhat better defined.  The improved image quality
aids in determining the angular size and frequency spectrum of various
components in a source.

\begin{figure}[ht]
\vskip 1cm
\includegraphics[angle=0,scale=0.4]{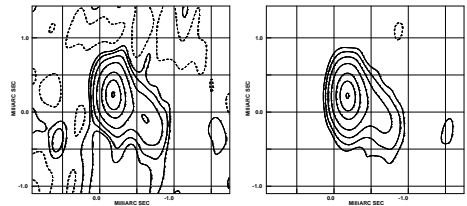}
\vspace{-1cm}
\caption{A comparison between the phase-reference (PR) image (left)
  and the self-calibrated (SC) image (right) for 0547 at 43 GHz on
  08-Jun-2009.  The SC image was obtained by starting with the PR
  image and then determining the residual antenna-based phase errors
  most consistent with the PR image.  Both images have been cleaned
  and the restoring beam for both is $0.6\times 0.3$ mas.  The contour
  levels are at -3,3,6,12,24,50,75,99.5\% of the peak brightness of
  0.087 Jy and 0.115 Jy for the PR and SC images, respectively.  The
  grid lines are separate by 0.5 mas.}
\label{fig:PR_SC_image}
\end{figure}

Tests from these VLBA and VERA observations (as well as from many
other VLBI sessions) conclusively show that the self-calibration
algorithm, when starting with the phase-referenced image, does not
change significantly the position of the radio emission in the
improved images.  In Fig.~\ref{fig:PR_SC_image}, the PR-image and the
SC-image are shown for 0547+234 at 43 GHz from the 08-Jun-2009
session.  This example was chosen because the separation between
0556+238 and 0547+234 is the largest of the pairs, about two degrees,
where residual phase errors can become as large as a radian at 43 GHz.
For this particular example, the difference in the location of the
peak brightness for PR and the SC image is 0.003 mas east/west and
0.015 mas north/south.  The position difference of the secondary peak
is 0.02 and 0.07 mas, well within the uncertainty of the location of
this faint and blended component.  For other source pairs and at 23
GHz frequencies, the differences between a PR and an SC image will be
less than that shown in this example.  Hence, in this paper, the
position of all components and the image displays will be obtained
from the self-calibrated data.

Several methods can be used to determine a \lq\lq position\rq\rq~ of a
component from a clean image including: 1) Measure the location of its
peak brightness.  This position determination is resolution dependent,
and is subject to the blending of close components.  This uncertainty
can be somewhat lessened if the same resolution is used for all
sessions, even though this degrades the resolution for some of them.
2) Fit the emission on an image with a small number of
Gaussian-components that are needed to describe the emission
properties.  Each component intensity, position and angular size, with
error estimates, are then obtained.  The angular size measured from
the image can be deconvolved (since the images resolution is known) in
order to obtain an estimate of the true sky angular size.  3) Fit the
calibrated visibility data (self-calibrated) directly with a small
number of Gaussian-components.  The initial guess often comes from a
clean image of the source.  Such modeling of the visibility data
directly, with no recourse to the image and its resolution limits, can
often determine a component's true angular size or more complex structure
more accurately than by image fitting techniques.

Method (2) was used for most of the component and source analysis.
When it is necessary to determine the internal structure of some of
the compact components, method (3) was used in order to obtain as high
a resolution as the data signal-to-noise would permit.

\subsection {The Morphology of Each Source}

\begin{figure}[ht]
\vspace{-10.3cm}
\hspace{-2.8cm}
\includegraphics[angle=0,scale=1.0]{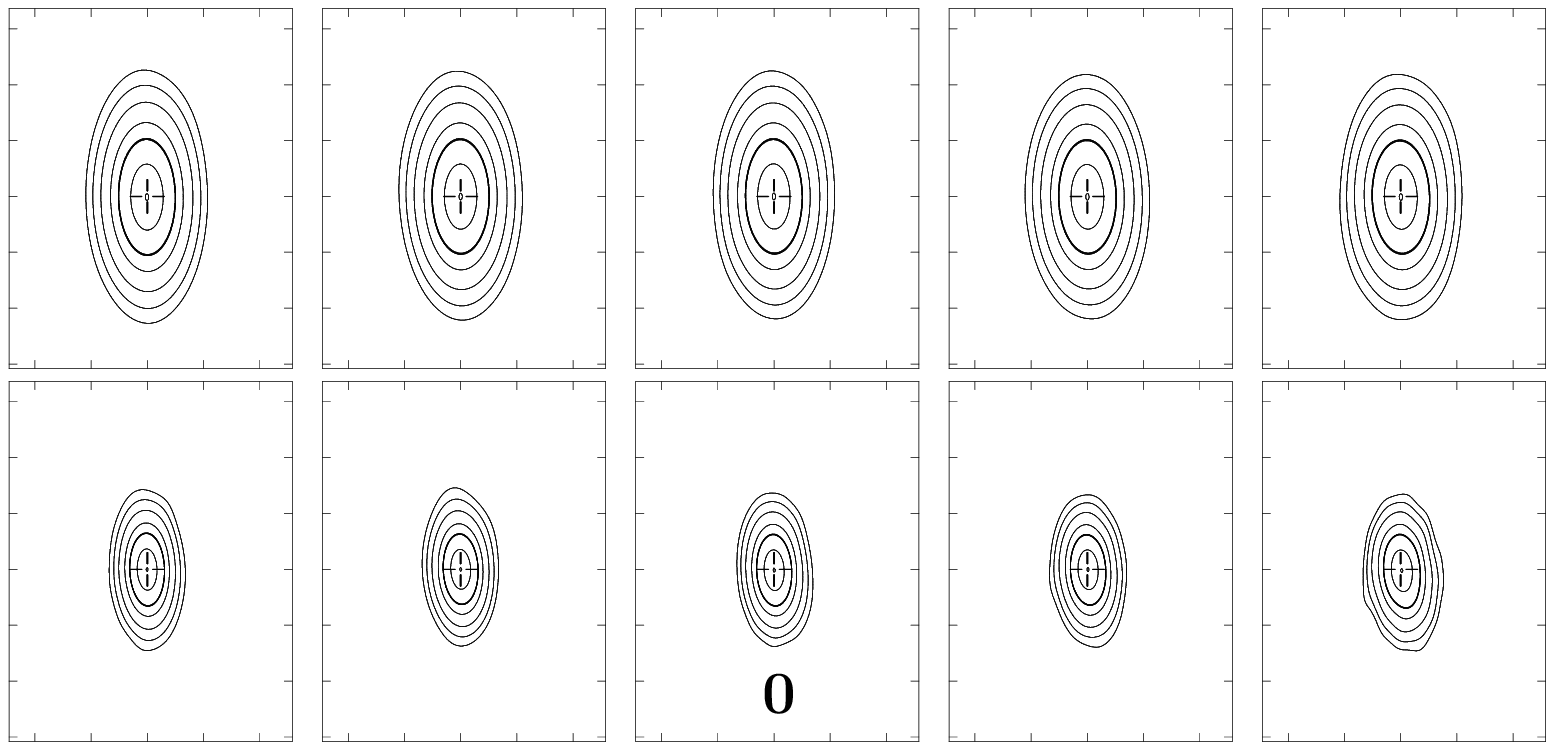}
\vspace{-9.5cm}
\caption{The VLBA images at 23 GHz and 43 GHz for the phase reference
  source 0556+238: The top row shows the 23 GHz contour plots from
  left to right for the sessions: 2008-Dec-21, 2009-Mar-16,
  2009-Jun-08, 2009-Sep-14 and 2009-Dec-09; respectively.  The bottom
  row shows the 43 GHz contour plots.  The resolution at 23 and 43 GHz
  is $1.0\times 0.5$ mas and $0.6\times 0.3$ mas, respectively.  The
  field of view for all images is $3.4\times 2.6$ mas, and the tick
  marks are separated by 0.5 mas.  The '+' shows the location of the
  phase center for the source.  All contour levels are at -4, 4, 8,
  16, 32, 50(bold), 80, 99.5 percent of the peak flux density.  This
  component is designated as {\bf 0}.}
\label{fig:f0556_images}
\end{figure}

\begin{figure}[ht]
\vspace{-10.3cm}
\hspace{-2.8cm}
\includegraphics[angle=0,scale=1.0]{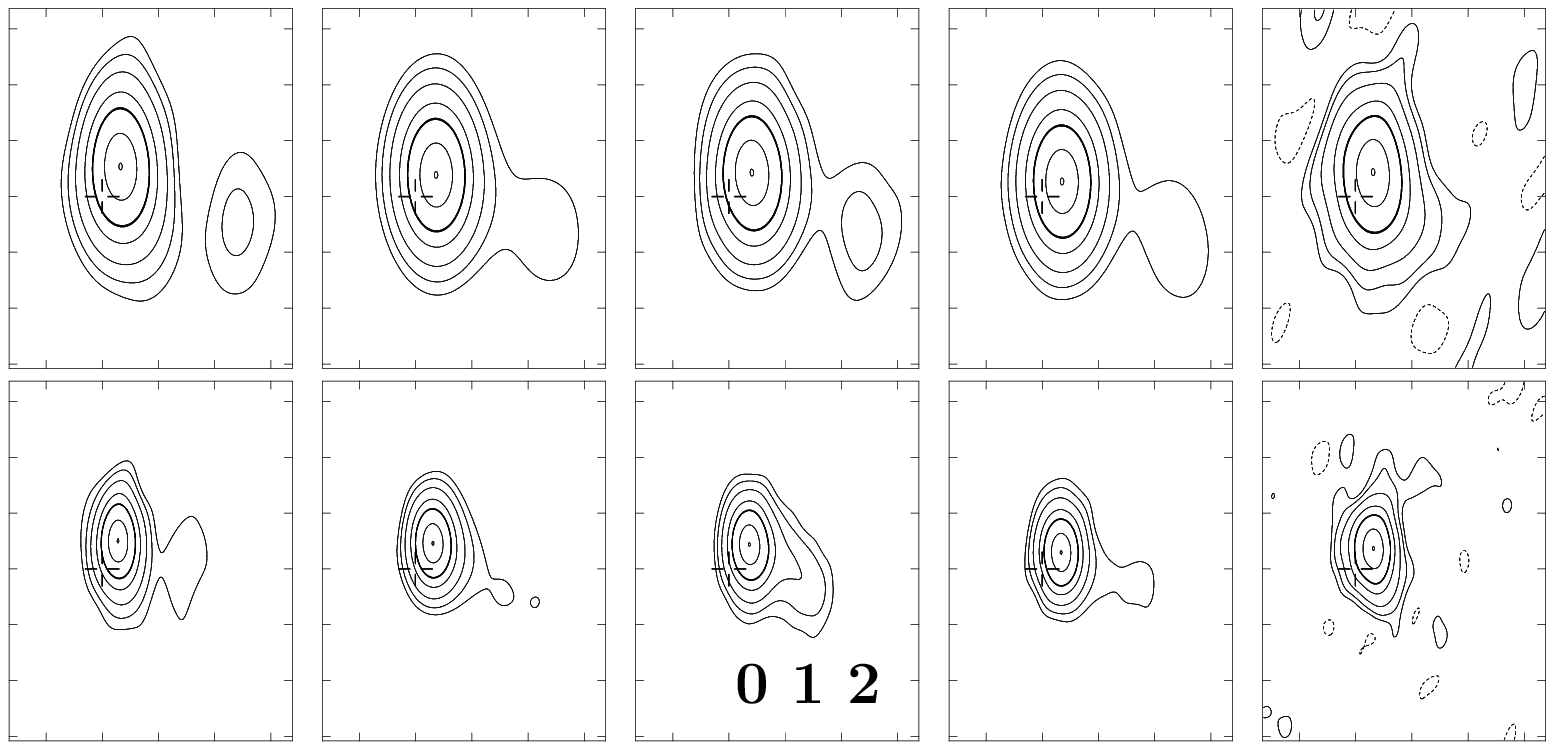}
\vspace{-9.5cm}
\caption{The VLBA images at 23 GHz and 43 GHz for the phase reference
  source 0547+234: The top row shows the 23 GHz contour plots from
  left to right for the sessions: 2008-Dec-21, 2009-Mar-16,
  2009-Jun-08, 2009-Sep-14 and 2009-Dec-09; respectively.  The bottom
  row shows the 43 GHz contour plots.  The resolution at 23 and 43 GHz
  is $1.0\times 0.5$ mas and $0.6\times 0.3$ mas, respectively.  The
  field of view for all images is $3.4\times 2.6$ mas, and the tick
  marks are separated by 0.5 mas.  The '+' shows the location of the
  phase center for the source.  All contour levels are at -4, 4, 8,
  16, 32, 50(bold), 80, 99.5 percent of the peak flux density.  The three
components contained in the source will be designated as {\bf 0, 1, 2}.}
\label{fig:f0547_images}
\end{figure}

\begin{figure}[ht]
\vspace{-10.3cm}
\hspace{-2.8cm}
\includegraphics[angle=0,scale=1.0]{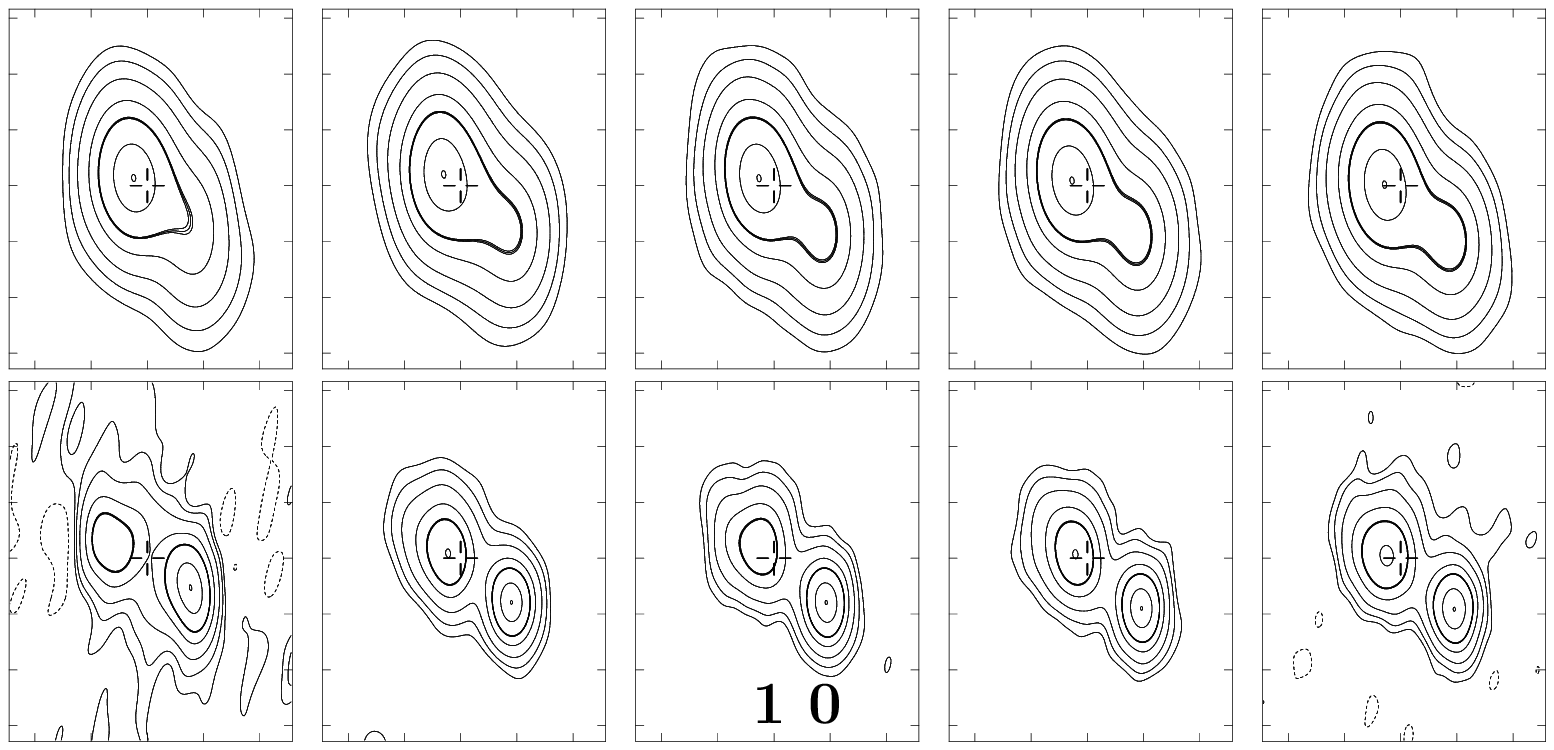}
\vspace{-9.5cm}
\caption{The VLBA images at 23 GHz and 43 GHz for the phase reference
  source 0554+242: The top row shows the 23 GHz contour plots from
  left to right for the sessions: 2008-Dec-21, 2009-Mar-16,
  2009-Jun-08, 2009-Sep-14 and 2009-Dec-09; respectively.  The bottom
  row shows the 43 GHz contour plots.  The resolution at 23 and 43 GHz
  is $1.0\times 0.5$ mas and $0.6\times 0.3$ mas, respectively.  The
  field of view for all images is $3.4\times 2.6$ mas, and the tick
  marks are separated by 0.5 mas.  The '+' shows the location of the
  phase center for the source.  All contour levels are at -4, 4, 8,
  16, 32, 50(bold), 80, 99.5 percent of the peak flux density.   The two
components contained in the source will be designated as {\bf 0, 1.}}
\label{fig:f0554_images}
\end{figure}

\begin{figure}[ht]
\vspace{-10.3cm}
\hspace{-2.8cm}
\includegraphics[angle=0,scale=1.0]{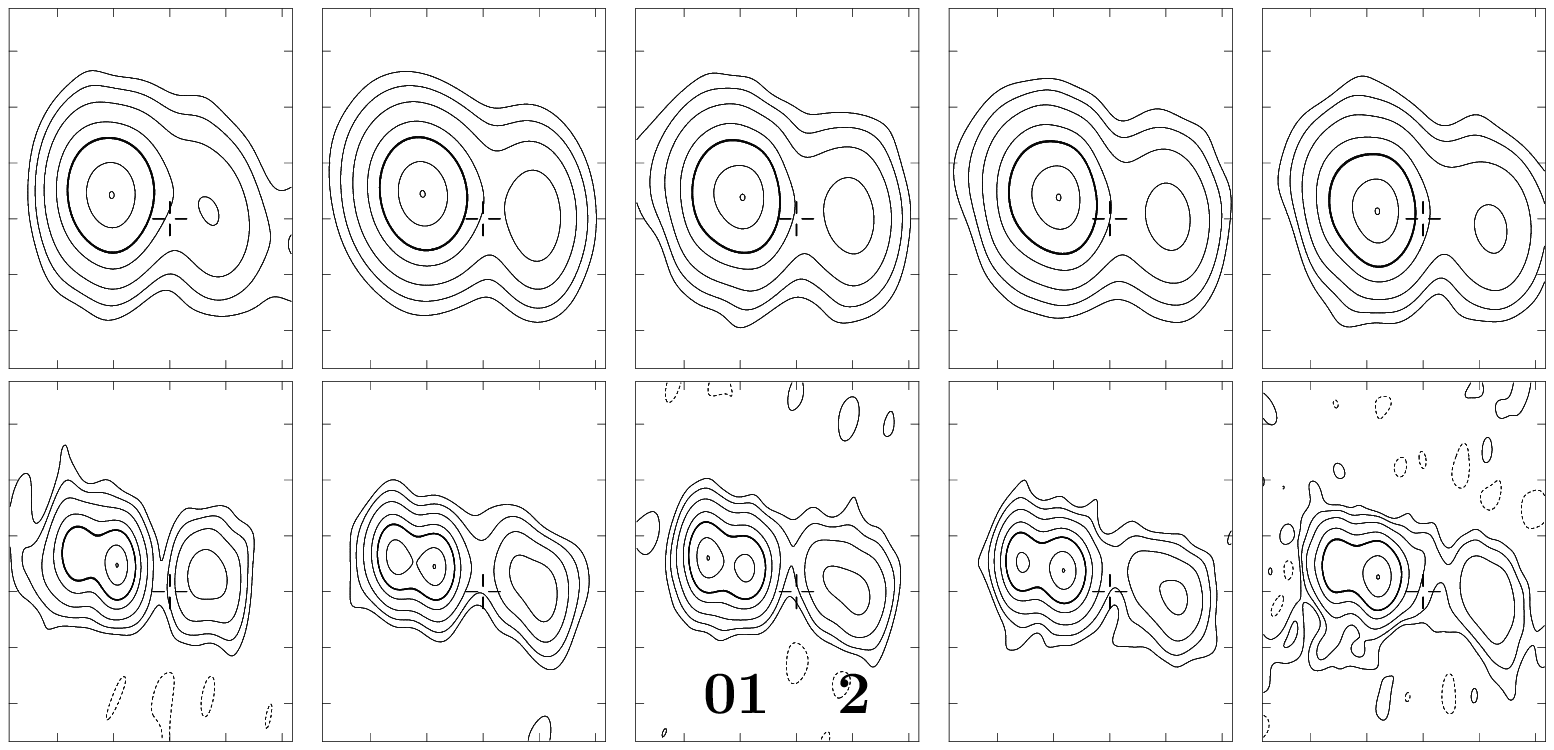}
\vspace{-9.5cm}
\caption{The VLBA images at 23 GHz and 43 GHz for the phase reference
  source 0601+245: The top row shows the 23 GHz contour plots from
  left to right for the sessions: 2008-Dec-21, 2009-Mar-16,
  2009-Jun-08, 2009-Sep-14 and 2009-Dec-09; respectively.  The bottom
  row shows the 43 GHz contour plots.  The resolution at 23 and 43 GHz
  is $1.0\times 0.5$ mas and $0.6\times 0.3$ mas, respectively.  The
  field of view for all images is $3.4\times 2.6$ mas, and the tick
  marks are separated by 0.5 mas.  The '+' shows the location of the
  phase center for the source.  All contour levels are at -4, 4, 8,
  16, 32, 50(bold), 80, 99.5 percent of the peak flux density.   The three
components contained in the source will be designated as {\bf 0, 1, 2}.}
\label{fig:f0601_images}
\end{figure}

The VLBA images in the form of contour images at 23 GHz and 43 GHz for
five sessions are given for the phase reference source, 0556+238, and
the three target sources, 0547+234, 0554+242 and 0601+245 in
Fig.~\ref{fig:f0556_images}, Fig.~\ref{fig:f0547_images},
Fig.~\ref{fig:f0554_images}, and Fig.~\ref{fig:f0601_images},
respectively.  Although there are six VLBA observing sessions, the first
contour plot on the left shows
the average of the images for the 20-Dec-2008 and
22-Dec-2008 sessions\footnote{The intent of these first two sessions,
  only two days apart, were to determine the approximate positional
  accuracy that could be obtained from these observations.  A lower
  sensitivity set-up was used and the 43 GHz data were not
  well-calibrated.  Improvement in the observing strategy were then
  made for the subsequent VLBA sessions.  The average of these two
  sessions are comparable in quality with the later sessions, so the
  results are included and are given the effective observing date of
  21-Dec-2008.}.  All images have the identical contour levels as a
percentage of the peak intensity.  The quality of the images at 43 GHz
on 12-Dec-2009 and 21-Dec-2008 (the average of two sessions) is
somewhat degraded because of inclement weather at several antennas.
This series of images show the good consistency of the VLBA images at
both frequencies, and that none of the sources changed substantially
over the year period.  The components for each source have been given
a number designation.  The component '0' is thought to be the core,
and will be discussed in detail in \S~\ref{sec:where_core}.

The images and structure for 0556+238 have been discussed in
\S~\ref{sec:0556_choice} in the decision to use this source as the
main phase reference for the experiments.  It is nearly a point source,
although the last session at 43 GHz shows a slight extension to the
south-west.  Deeper imaging of this sources will show the surprising
evolution of this source structure over the year observation in \S
\ref{sec:super_resolution}.

The source 0547+234 is dominated by a compact core, displaced about
0.5 mas north-west of the phase center, shown by the '+'.  A faint
component about 1.0 mas to the west is somewhat variable and brightest
on the 08-Jun-2009 session.  The source 0554+242 has two components,
clearly shown in the 43 GHz images.  The phase center is located near
the component to the north-east, but the south-west component appears
more compact and is possibly the radio core.  Finally, the source
0601+245 has a complex structure that is resolved into three major
components at 43 GHz.  The possible location of the radio core in this
source is unclear with this level of display.

\subsection {The Frequency Dependence and \\ VLBA/VERA Comparisons}

Most of the analysis in the paper will be derived from the VLBA
observations since these were the highest resolution and covered the
frequency range from 8.6 GHz to 43 GHz.  The most accurate VERA
session was on 17-Jan-2009 when the weather at the four antennas was
excellent.  For the other VERA sessions one antenna was not available
for a substantial part of the 6.5-hour observation, primarily because
of inclement weather.

\begin{figure}[ht]
\vspace{-4.5cm}
\hspace{0cm}
\includegraphics[angle=0,scale=0.8]{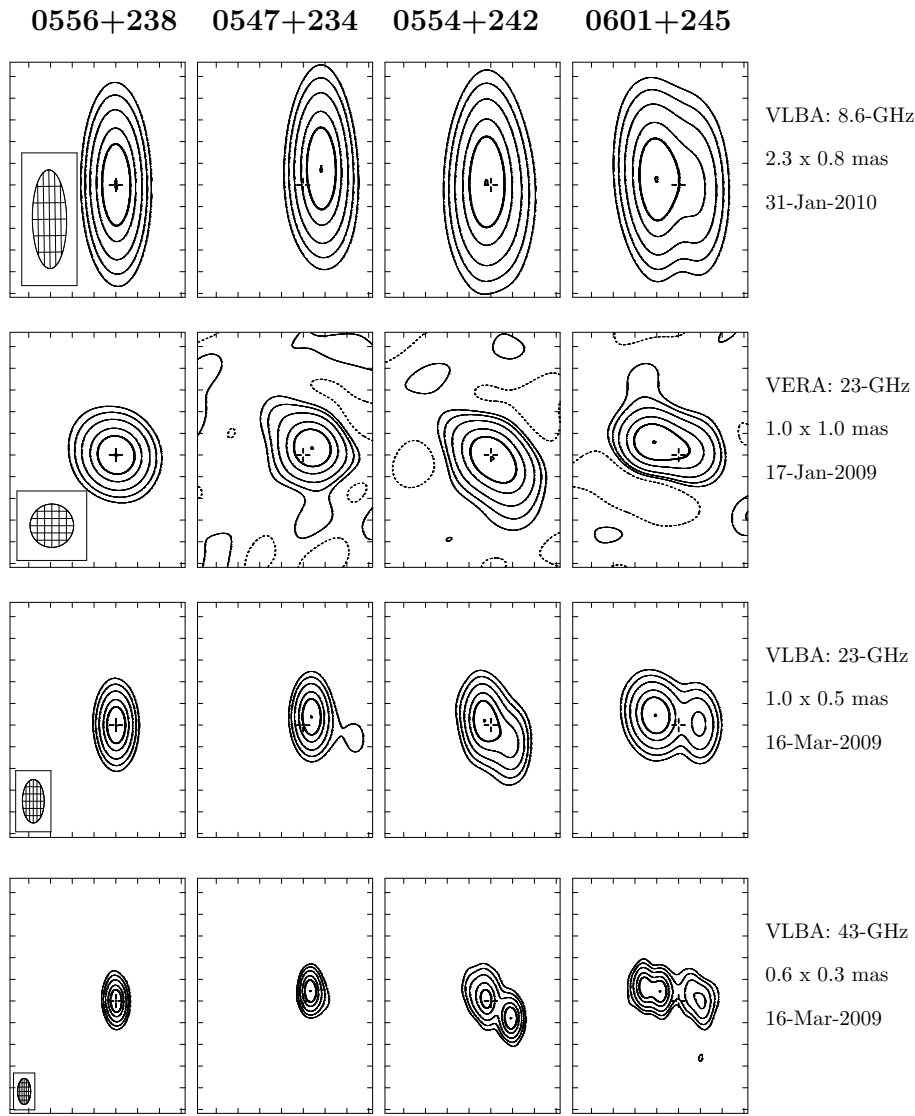}
\vspace{-3.5cm}
\caption{The contour images at several frequencies for the VLBA and
  the VERA comparison: The images for the four sources are given in
  each column.  At the end of each row the array, frequency,
  resolution, and session date are listed.  The resolution for all
  sources is shown by the cross-hatched ellipse for 0556+238.  The
  field of view for all images is $5.4\times 4.0$ mas, and the
  tick-mark are at 0.5-mas intervals.  The '+' shows the location of
  the phase center assumed for each source.  All contour levels are at
  -5, 5, 10, 20, 50(bold), 75, 99.5 percent of the peak flux density.}
\label{fig:freq_comparison}
\end{figure}

The comparison among the VLBA images at 8.6, 23 and 43 GHz with the
VERA images at 23 GHz are shown in Fig.~\ref{fig:freq_comparison}.
The importance of high resolution is demonstrated in these
comparisons.  For example, the VERA and VLBA 23 GHz images show
similar structures and position for the four sources, but some
essential features of the source structure are more apparent with the
better VLBA resolution, especially at 43 GHz.  The VERA images,
although not used heavily in the following analysis, were consistent
in structure and position with the VLBA images, lending more
confidence to the overall conclusions.  The non-contemporaneous aspect
of the images shown in this figure are not a complicating factor in
their comparison because of the relative slow pace of evolution for
the sources.

The VLBA resolution at 8.6 GHz, the same frequency used for the ICRF,
barely hints at the inner structure of these four sources.  Clearly,
without the 43 GHz VLBA images at the highest resolution,
interpretation of the source structure and evolution would be
ambiguous---as will be documented in later sections.

\section {Analysis of the Radio Source Structure and Evolution}

\subsection {Finding the Radio Core in each Source} \label {sec:where_core}

The major goals of the observations of these four sources were 1) to
determine the location of the radio core for each source, 2) to
determine their position stability, and 3) to assess the affect of
structure on the ICRF2 positions.  Finding the core components for
each of the four sources involved somewhat different combination of
most of the following properties: (1) The most compact component with
0.3-mas resolution; (2) The component with the flattest spectral
index; (3) The component at one end of the emission region; (4) The
component that is stationary with respect to the radio cores in other
sources.  The use of these four criteria will become clear in the
following discussions.

The four sources that have been chosen, although relatively compact,
according to the ICRF2 catalog, have significantly different
structural properties with sub-mas resolution imaging.  The two
sources that appear to be core-dominated are 0547+234 and 0556+238,
which was used as the phase calibrator.  On the other hand, the other
two sources have sufficiently complex structure so that the location
of the radio core within the emission is not obvious from the images
in Fig.~\ref{fig:f0554_images} and Fig.~\ref{fig:f0601_images}.  The
multi-frequency, multi-session observations of these two sources were
needed to \lq\lq find\rq\rq~their radio core.

\subsection{The Structure and Evolution \\ of 0556+238}

The source 0556+238 is nearly point-like at all frequencies and
sessions, which is the main reason that it was used as the phase
reference calibrator.  The last session at 43 GHz shows a faint
extension of the emission to the south-west.  The VLBA images were fit
to a single elliptical Gaussian component, labeled {\bf 0} in
Fig.~\ref{fig:f0556_images}, and its parameters are listed in
Table~\ref{tab:0556}.  Column 1 gives the session data and the next
five columns give the component total flux density, peak flux density
and the estimated error of the peak, and the offset of the component X
(east) and Y (north) with error estimates, all at 43 GHz.  The next
five columns give the same intensity and position estimates at 23 GHz.
The ratio of the total flux density to the peak flux density, slightly
above unity, can be caused by the remaining small phase errors
(decorrelation) or from a non-zero angular size of the component which
corresponds to a full width at half-power diameter of about 0.2 mas.
Over the experimental period of about one year, the flux density of
the radio core increased 46\% at 43 GHz and 18\% at 23 GHz, with the
spectral index, $\alpha$ (where flux density $\propto \nu^\alpha$),
changing from $-1.2$ to $-0.8$.

\subsection{The Structure and Evolution \\ of 0547+234}

The source 0547+234 is dominated by a compact component toward the
east.  At 43 GHz, the source structure was fit with an elliptical
Gaussian component for component {\bf 0} (clearly the radio core), and
with a point component for component {\bf 1} (it is too weak for an
angular size to be determined).  At 23 GHz the bright component radio
core and the fainter component {\bf 1} were each fit with point
source.  An additional component {\bf 2}, located about 1.2 mas to the
west, was added to the source model in order to fit the more extended
emission that is seen in Fig.~\ref{fig:f0547_images}.  The parameters
associated with components {\bf 1} and {\bf 2} are relatively
inaccurate since they are not particularly discrete features and
appear to change quickly with time, but are included in the fit in
order to remove any interaction of their parameters with that of the
radio core.  These parameters have not been included Table
\ref{tab:0547}.  Observations of the source at 23 GHz in Aug 2005 show
a similar structure with a dominant core and a faint component about 1
mas to the east \citep{kq-survey2}.

The parameters for the radio core for the five VLBA sessions at 43 GHz
and 23 GHz are shown in Table~\ref{tab:0547}, with the same format as
that for Table~\ref{tab:0556}.  The component {\bf 0} flux density
reaches a minimum for the 08-Jun-2009 session at both 43 and 23 GHz,
when components {\bf 1} and {\bf 2} are near their maximum, about 15\%
of the flux density of the radio core, and the component spectral
index is about -0.6.  It is offset from the phase center by about
X=-0.18 and Y=+0.21 mas.  The most surprising property of component
{\bf 0} is its proper motion to the south at 23 GHz and 43 GHz,
approximately $\mu_y=0.12$ mas/yr.!  Slight motion to the west may
also occur.  The cause of this apparent motion is discussed below.

\subsection{The Structure and Evolution \\ of 0554+242}

The contour plots of 0554+242 in Fig.~\ref{fig:f0554_images} show two
major components.  The emission was fit with a two-component elliptical
Gaussian model for all sessions and at 43 and 23 GHz, and the
parameters are shown in Table~\ref{tab:0554}.  Component {\bf 0} is
only slightly resolved since the total to peak flux density is about
1.1, corresponding to an angular size $<0.2$ mas.  There is no
significant variability over the year experiment.  The X-position of
this component shows a small motion to the west, but, as with
0547+234, this component is moving to the south by about 0.09 mas over
the year.

Component 1 is clearly extended with the total flux density to peak
flux density ratio about 1.9 at 43 GHz and 1.3 at 23 GHz.  This ratio
corresponds to an angular size of about 0.8 mas at both frequencies.
The spectral index of the two components are significantly different.
For component {\bf 0}, $\alpha=-0.7$; for component {\bf 1},
$\alpha=-1.2$.  It is, thus, extremely likely that the radio core of
0554+242 is component {\bf 0} because it is more compact and has a
flatter spectrum than component {\bf 1}.

The relative proper motion between the two components (component {\bf
  1} - component {\bf 0} is $\mu_x=+0.018\pm 0.012$ mas/yr and
$\mu_y=+0.010\pm 0.015$ mas/yr, or $\mu=0.020\pm 0.013$ mas/yr in
position angle $56^\circ$; but it is only marginally detected.  The
component separation in June 2009 is 0.782 mas in position angle
$50^\circ$ which is fortuitously in the same direction as the possible
motion.  Observations at 23 GHz with the VLBA in 2005 August show a
separation of the two components of 0.75 mas in position angle
$54^\circ$ with flux densities 0.29 (NE) and 0.10 (SW-core)
\citep{kq-survey2}.  Earlier observations at 8.6 GHz from the 1990's
also suggest a similar separation of two major components.  Thus, the
two components are barely if at all separating, with an ejection time
that was at least 20 years in the past.

\subsection {The Structure and Evolution of \\ 0601+245}

The 43 GHz images of 0601+245 suggest that the source is composed of
three major components.  There are two close components in the eastern
part of the source and an extended component to the west (see
Fig.~\ref{fig:f0601_images}).  The location of a radio core, perhaps
corresponding to one of these components, cannot be easily deduced
from this display without deeper analysis.

Table~\ref{tab:0601} lists the parameters for the three components.
The two components to the east were modeled as point sources (when two
components overlap, determination of their individual angular sizes is
not reliable), but the western component was fit to an elliptical
Gaussian component, and it is clearly extended.  The component flux
densities are relatively stable at 43 GHz, but appear to be slowly
decreasing at 23 GHz.  Component {\bf 0} has the flattest spectral
index with with $\alpha=-0.7$, component {\bf 1} has a steep index of
$\alpha=-1.7$, and the western, extended component {\bf 2} has
$\alpha=-0.8$.

Component {\bf 0} shows a similar proper motion as the radio cores in
0547+234 and 0554+242 with $\mu_x=-0.029\pm 0.015$ mas/yr,
$\mu_y=-0.064\pm 0.025$ mas/yr.  Again, this anomalous proper motion
is discussed in the next sections.  This component is likely the radio
core since it also has the flattest spectral index, is compact, and is
at one end of the radio emission.

Component {\bf 2} is moving with $\mu_x=-0.183\pm 0.025$ mas/yr,
$\mu_y=-0.064\pm 0.033$ mas/yr, or $\mu=0.193\pm 0.029$ mas/yr in position angle
$-109^\circ$.  The separation of component {\bf 2} from {\bf 0} is
$X=-1.231\pm 0.022$, $Y=0.236\pm 0.040$, 1.253 mas in position angle
$-101^\circ$, which extrapolates to an ejection time of component {\bf
  2} from component {\bf 0} about 6.5 years ago.  Observations at 23
GHz with the VLBA in 2005 August shows the presence of only one
component, which is consistent with these measurements \citep{kq-survey2}.

The motion of component {\bf 1} away from {\bf 0} is $\mu_x=-0.036\pm
0.020$ mas/yr, $\mu_y=-0.029\pm 0.030$ mas/yr, and is formally not
detected.  Although the separation is in the sense that component {\bf
  1} is moving away from {\bf 0}, the blending of the two components,
even at 43 GHz, makes determining their separation uncertain.

\section {Anomalous Position Behavior \\ for 0556+238}

\subsection {Consistency of Target Core Apparent Proper Motions}

The radio cores in 0547+234, 0554+242 and 0601+245 have been
tentatively identified.  For 0547+234 this association was
straight-forward since the emission is dominated by a single
component.  For 0554+242 and 0601+245 the 23 GHz and 43 GHz high
resolution images were both needed to determine the core location: it
was the most compact component, and with the flattest spectral index.
What is surprising is that cores in the three target sources have
nearly the same proper motion, mainly to the south!

\begin{figure}[ht]
\vspace{-0.6cm}
\hspace{0cm}
\includegraphics[angle=270,scale=0.8]{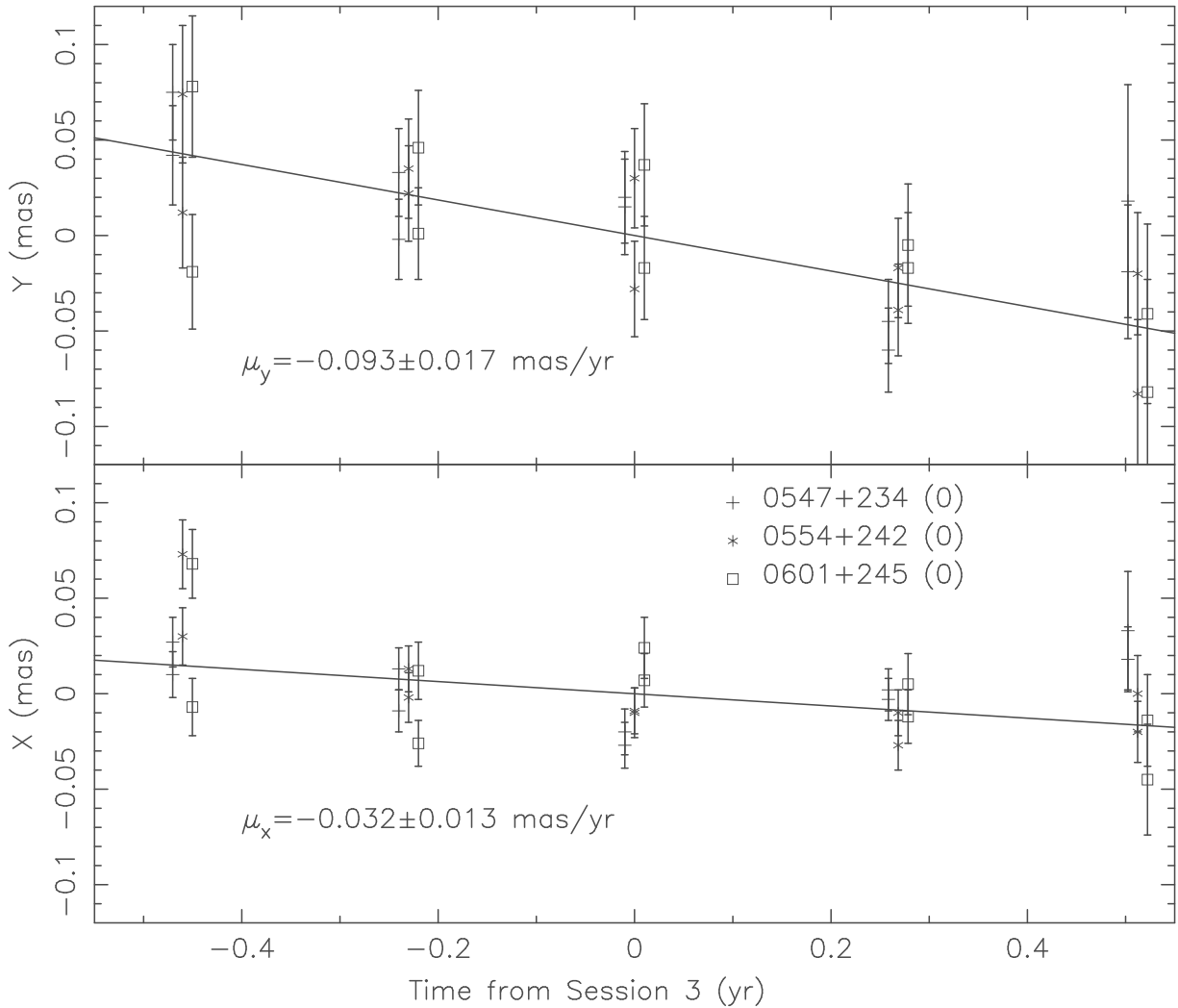}
\vspace{0.5cm}
\caption{The Motion of the Cores of 0547+234, 0554+242 and 0601+245:
  The position of the core radio components, after removal of the
  position offset, are superposed for the three sources.  Both 43 GHz
  and 23 GHz points are plotted and the source data is indicated by
  the symbol.  Each source point is slightly displaced in time in
  order to distinguish among the sources.  The data and error
  estimates comes directions from the tables of the source parameters.
  The average proper motion in X and Y are shown by the straight
  lines.}
\label{fig:all_motion}
\end{figure}

To investigate these proper motions in more detail, the positions
measured for the core components from
Tables~\ref{tab:0547},\ref{tab:0554} and \ref{tab:0601} are plotted in
Fig.~\ref{fig:all_motion}.  Both the 43 GHz and 23 GHz core positions
were used, although there could be a small separation between them,
but this should not bias the proper motion fit.  The relative offset
of the cores from their phase center has been removed, so the motions
of the three sources could be overlapped.  These core offsets were:
0547+234 (X=-0.171, Y=+0.194) mas; 0554+242 (X=-0.454, Y=-0.429) mas;
0601+245 (X=+0.772, Y=+0.270) mas.  The first session and the last
session points show larger scatter than the middle three sessions, as
suggested from the contour plots for these two sessions in
Fig.~\ref{fig:f0556_images}-\ref{fig:f0601_images}.

The proper motion fit to the core positions of the three target
sources are $\mu_x=-0.032\pm 0.013$ mas/yr, $\mu_y=-0.093\pm 0.017$
mas/yr, and all three cores are consistent with this motion.  Thus,
the relative separation among them remained constant to the
measurement error of about 0.02 mas, as indicated by the scatter of
the points around the best fit proper motion line in
Fig.~\ref{fig:all_motion}.  The similar core motion for the three
target sources strongly implies that the emission peak of 0556+238,
which defines the location of its apriori position, is moving with
respect to the other three cores.  In the next section,
super-resolution imaging and modeling of the emission from 0556+238,
will show how internal changes in this source structure have mimicked
a proper motion during the experimental period.

\subsection {Super-resolution Imaging of 0556+238} \label{sec:super_resolution}

Although 0556+238 is nearly a point source with the nominal
resolutions at 23 GHz and 43 GHz, the source intensity does vary
significantly, with a change in spectral index, over the year
experiment.  Hence, there is some indication of internal changes, even
within the resolution of the 43 GHz observations.  The source has been
well-observed for decades in the ICRF-programs with little indication
of extended structure.  The derived positions from the hundreds of
ICRF experiments are stable at the 0.2 mas level, although this is the
typical position error for any one ICRF session.  It has a structure
index of 1, meaning that it has no significant structure outside of
the core.  It is one of the 295 ICRF2 sources used to define the
celestial reference frame.

Errors in the data processing and analysis that could produce the
apparent motion of 0556+238 are unlikely.  Unexpected astrometric and
tropospheric errors should produce errors that are functions of the
calibrator target separation-and orientation, rather than a constant
offset for all targets.  An error in the data processing and the
correlator model in which an anomalous proper motion was inserted in
the position of 0556+238 over the year experiment has not occurred.

\begin{figure}[ht]
\vspace{-15.0cm}
\hspace{-5.5cm}
\includegraphics[angle=0,scale=1.4]{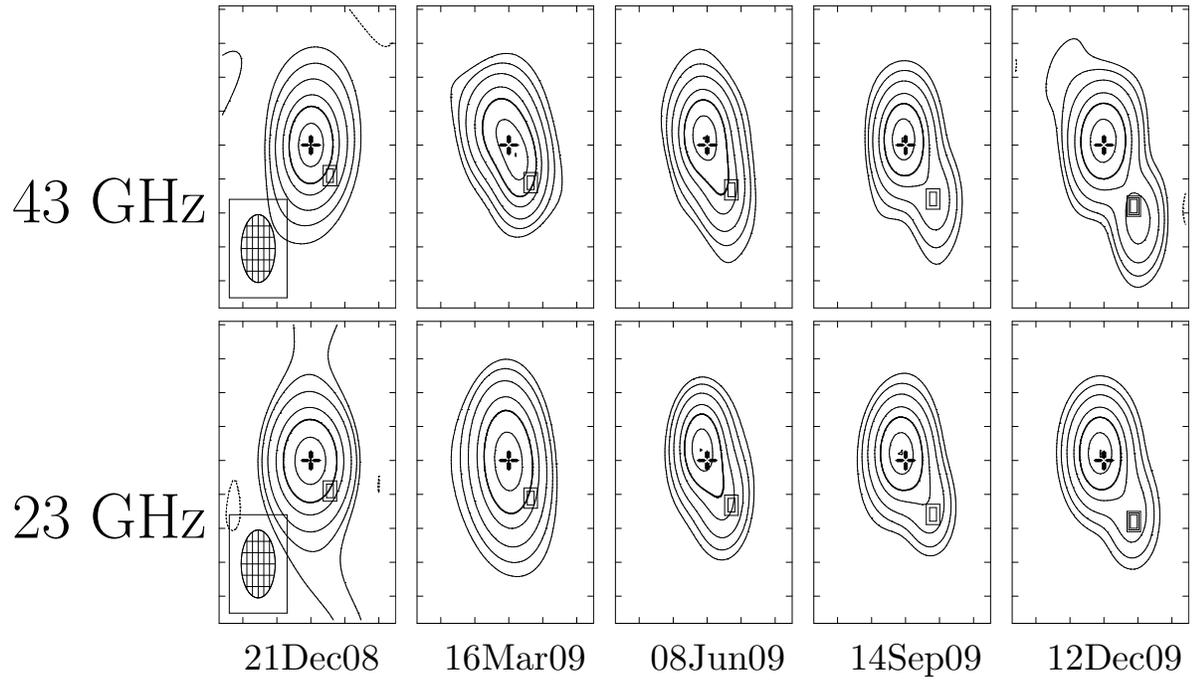}
\vspace{-15.0cm}
\caption{Super-resolution images of 0556+238: Images for the five
  sessions are shown at 43 GHz and 23 GHz.  The assumed phase center
  is show by the '+'.  A stationary point in this source, derived
  from the apparent core motions of the three target sources, is shown
  by the 'box'.  The restoring beam at both frequencies, shown by the
  cross-hatched ellipse in the left-most diagram of each row, is
  $0.2\times (N/S) 0.1$ mas.  The contour levels are at -4, 4, 8, 16, 32,
  50(bold), 80, 99.5 percent of the peak flux density.  The tick marks
  are separated by 0.1 mas.}
\label{fig:0556_motion}
\end{figure}

The 43 GHz and 23 GHz data for 0556+238, after editing and
self-calibration in AIPS, were transferred to the Caltech
interferometric package, Difmap \citep{DIFMAP}.  The visibility data
were then fit with one or two Gaussian components.  For the first two
sessions when the source was small in angular size, only one extended
Gaussian was fit; but for the three final sessions, when the source
became more extended, a double point model was used.  These parameters
are shown in Table~\ref{tab:0556_motion} and the images, with a
restoring beam of $0.2\times 0.1$ mas at both frequencies, are
displayed in Fig.~\ref{fig:0556_motion}.  This resolution is about a
factor of 2 smaller than the conservative one used for previous image
display at 43 GHz and a factor of 3 at 23 GHz in
Fig.~\ref{fig:f0556_images}.

The NE component is the brighter than the SW component by a factor of
three at 43 GHz and five at 23 GHz, with the spectral index of the NE
and SW components $\alpha\approx -0.8$ and $\alpha\approx 0.0$,
respectively.  At both frequencies the total flux density of the
source at both frequencies increased by about 50\% from Dec 2008 to
Dec 2009.  The '+' in the contour images of 0556+238 in
Fig.~\ref{fig:0556_motion} shows the location of the assumed apriori
position for the source.  The location is determined by the
self-calibration algorithm and is usually within 0.005 mas of the peak
brightness.  Since the NE component is much brighter than the SW
component, the apriori phase center is virtually coincident with the
NE component.

The source contains two components that are moving apart.  From the
model fits the component separations are 0.15, 0.19, 0.24 mas for the
June, Sep and Dec 2009 sessions, with good agreement between the 23
and the 43 GHz results.  Using the ratio of peak to integrated flux
densities for the first two sessions as a measure of the separation of
the two components, we obtain about 0.10 mas for both sessions.  Thus,
to the accuracy of these fits, the proper motion between the two
components is about $0.14\pm 0.02$ mas/yr in position angle $150\pm
15^\circ$.  For comparison, the average proper motion of the three
cores in the target sources (See Fig.~\ref{fig:all_motion}) is $0.10\pm
0.02$ mas/yr in position angle $161\pm 12^\circ$.  If these three
cores are assumed to be stationary in the sky, then a stationary point
in the images for 0556+238 over the sessions is shown by the 'box'.
It is clear that the SW component is not moving significantly with
respect to the other three target radio cores.

The conclusion is that the south-west component is the {\it
  stationary} radio core of 0556+238.  The brighter component is a new
jet component, moving to the north-north east with a proper motion of
about 0.12 mas/yr.  It appears fortuitous that this motion is
quasi-linear because the first session occurred somewhat after the jet
ejection, and the jet component velocity has remained relatively
constant over the year period, a typical behavior of many jet
components just after their formation \citep{lis09}.

\subsection {Super-resolution Imaging of Other Sources}

The visibility data were also analyzed for the other sources, using
similar super-resolution techniques that were described above.  The
cores for the three target sources were compact for all of the
sessions, and were sufficiently far from the additional extended
emission, so that their position could be unambiguously determined to
an accuracy of 0.02 mas.  The unusual circumstances affecting the
interpretation of the structure of 0556+238---the ejection of a new
component from the core during the experiment period, and the
intensity of this component significantly greater than that of the
core---did not occur for the other three sources.

\section {Comparison with the ICRF2 and the 8.6 GHz Positions}

The Jan 2010 VLBA session observed the four sources in phase-reference
mode at 8.6 GHz for two main reasons: First, to obtain the 8.6 GHz
structures and positions that are relatively co-temporaneous to the 23
and 43 GHz images in Dec 2009, so that time variability is not a
serious problem for their comparison.  Second, to assess the ICRF2
positions with the detailed structure information obtained for the
four sources.

\subsection {ICRF2 Registration of Images using 0556+238}

\begin{figure}[ht]
\vspace{-6cm}
\hspace{-2cm}
\includegraphics[angle=0,scale=0.9]{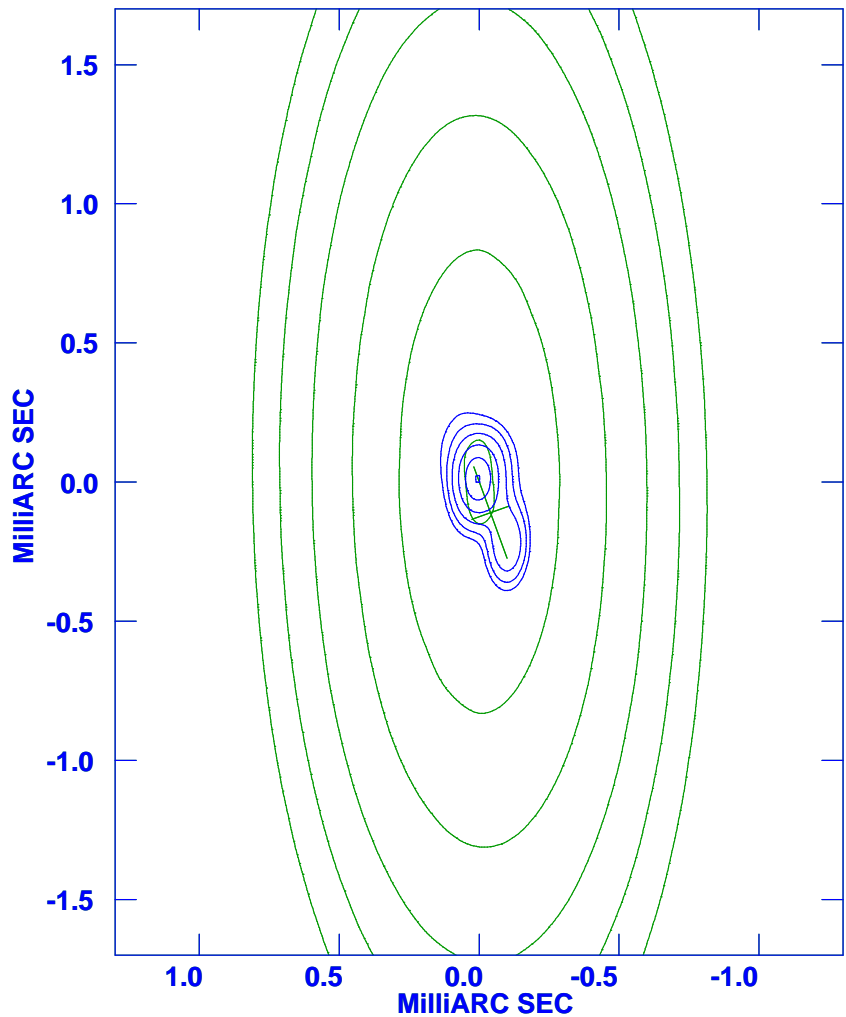}
\vspace{-5cm}
\caption{Frequency-Dependent Structure of 0556+238: The green contours
  show the Jan 2010 8.6-GHz VLBA phase-referencing image.  The blue
  contours show the Dec 2009 43 GHz VLBA phase-referencing image.  The
  green error bar indicates the best estimate of location of the ICRF2
  position (see text).  The contour levels for both images are at 5,
  10, 25, 50, 75, 99\% of the peak.  The tick marks are separated by
  0.5 mas.}
\label{fig:0556_registration}
\end{figure}

An overlay of 0556+238 of the 43-GHz image in Dec 2009 and the 8.6-GHz
in Jan 2010 image is shown in Fig.~\ref{fig:0556_registration}.  The
observations were taken 45 days apart, so the slight variability or
motion in the source should not be a contributing factor in the image
differences.  

However, the 43 GHz and 8.6 GHz grid frames could be displaced.  This
is unlikely because the assumed phase center (same for all three
frequencies) is virtually located at the position of the peak
brightness, and the NE component is much brighter than the SW
component at 8.6, 23 and 43 GHz.  Thus, the Dec 2009 23 and 43 GHz
images and Jan 2010 8.6 GHz images are on the same reference grid to
better than 0.1 mas.  This means that the comparison of the 8.6 GHz
and 43 GHz images for the other sources will also be accurately
registered since these are all tied to that of 0556+238, the phase
reference calibrator.

\subsection {Where is the ICRF2 Position of 0556+238?}

The ICRF2 position of 0556+238, one of the 295 sources that are used
to define the ICRF quasi-inertial reference frame, is
$\alpha=05^h59^m32.03313165^s, \delta=23^\circ 53'53.9267683''$ with
an estimated error of 0.04 mas in both coordinates.  This improved
position is 0.026 mas east and 0.068 mas north of the assumed phase
center (see Table~\ref{tab:source_parameters}), that was taken from
the previous ICRF catalog, and the ICRF2 position is 0.07 mas
displaced in the same direction as that between the NE and SW
components!

An estimate of the evolution of the structure of this source structure
can be made from VLBA observations at 23 and 43 GHz in Dec 2002, Sep
2003 and from 23 GHz and 8.6 GHz from Aug 2005 \citep{kq-survey1,
  kq-survey2}.  Over-resolved images of the source for these
observations show two components that were separated between 0.1 and
0.3 mas in a position angle roughly $-135^\circ$.  The relative flux
density between the two components was not fixed, but the southern
component had a flatter spectral index in two of the three
experiments, which supports the contention that the radio core is
indeed the south-west component.  The time-scale of the formation and
motion this jet component cannot be ascertained.

The location of the ICRF2 position of 0556+238 could lie anywhere
between the NE and SW components, although at 8.6 GHz, the NE
component is probably the stronger of the two components and the ICRF2
position may favor this position.  The error cross in
Fig.~\ref{fig:0556_registration} is a conservative estimate of the range of
the location of the ICRF2 position in Dec 2009 to Jan 2010.  This error
box will be transferred to the target sources in order to compare their
23 and 43 GHz images with the 8.6 GHz images and with the ICRF2 position.

\subsection {0547+234 at 8.6 GHz}

\begin{figure}[ht]
\vspace{-6cm}
\hspace{-2cm}
\includegraphics[angle=0,scale=0.9]{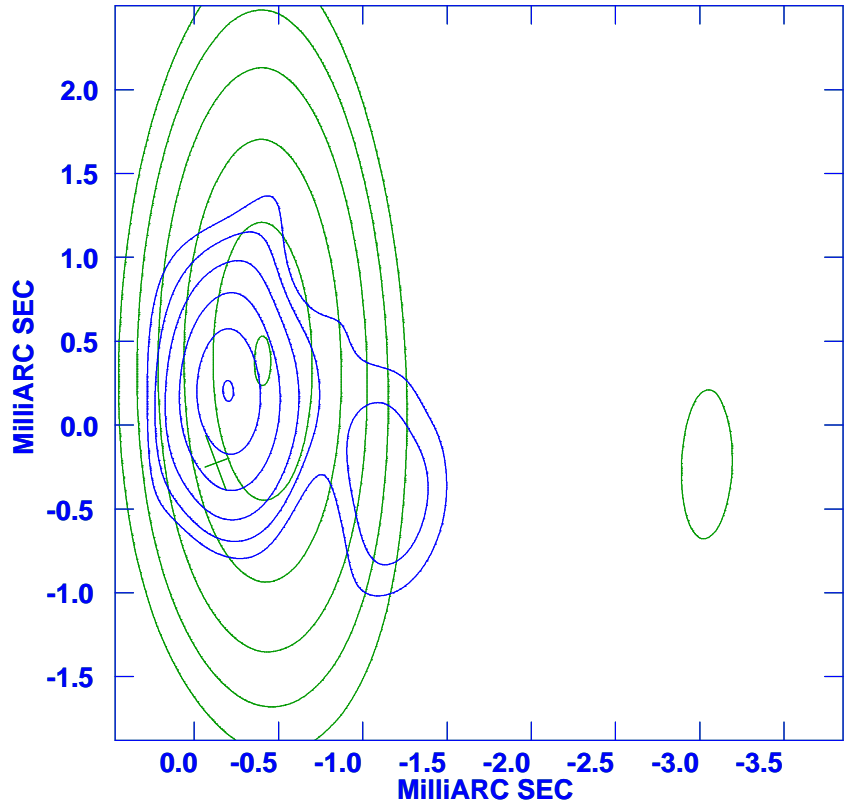}
\vspace{-5cm}
\caption{Frequency-Dependent Structure of 0547+234: The green contours
  show the Jan 2010 8.6-GHz VLBA phase-referencing image.  The blue
  contours show the June 2009 23 GHz VLBA phase-referencing image.
  The green error bar gives the estimate of the location of the ICRF2
  position for this source, derived from the registration of the
  0556+238 images.  The contour levels for the 8.6 GHz image are at
  -1,1,2,4,8,16,32, 50, 75, 99\% of the peak.  The contour levels for
  the 23 GHz image are at -5,5,10,20,40,50,75,99\% of the peak.  The
  tick marks are separated by 0.5 mas.}
\label{fig:0547_registration}
\end{figure}

A comparison of the 23 GHz Jun 2009 and and 8.6 GHz Jan 2010 VLBA
images is given in Fig.~\ref{fig:0547_registration}.  The June 2010
session at 23 GHz was used since it shows best the extended emission
to the west.  (The motion of the radio core between June and Dec 2009
was less than 0.1 mas which does not affect the registration of the
two images.)  Two anomalies are apparent.  First the ICRF2 position is
significantly south of the peak emission at all frequencies.  This
source has been observed in only about 23 ICRF sessions over the
years, and the formal ICRF2 position error is 0.3 mas.  Hence, the
offset of the ICRF2 position could be consistent with the images.

The second anomaly, the displacement of the peak brightness at 8.6 GHz
and 43 GHz (and all of the high-frequency images), is significant.  As
seen in Fig.~\ref{fig:f0547_images}, the faint emission to the west
comes and goes over 2009, at both 23 and 43 GHz.  A radio image at 8.6
GHz in Aug 2005 shows even more extended structure to the west of the
peak (http://rorf.usno.navy.mil/RRFID/) than that in Jan 2010.  An
explanation for this discrepancy is that there is significant diffuse
and variable radio emission emanating north and west of the radio core.
This emission may be optically thick within 1 mas of the core and, is
thus, barely detected at 23 and 43 GHz.  Thus, the change of position
of the core between 43 GHz and 23 GHz may be the so-called frequency
dependent core-shifts, see for some other radio sources \citep{kov09},
which is typically about 0.2 mas between 23 GHz and 8.6 GHz.  This is
the only one of the four sources which has extended emission near the
radio core.  Nearly all of the emission in the other three sources are
confined to discrete components and such core-shifts are not apparent.

\subsection {0554+242 at 8.6 GHz}

\begin{figure}[ht]
\vspace{-6cm}
\hspace{-2cm}
\includegraphics[angle=0,scale=0.9]{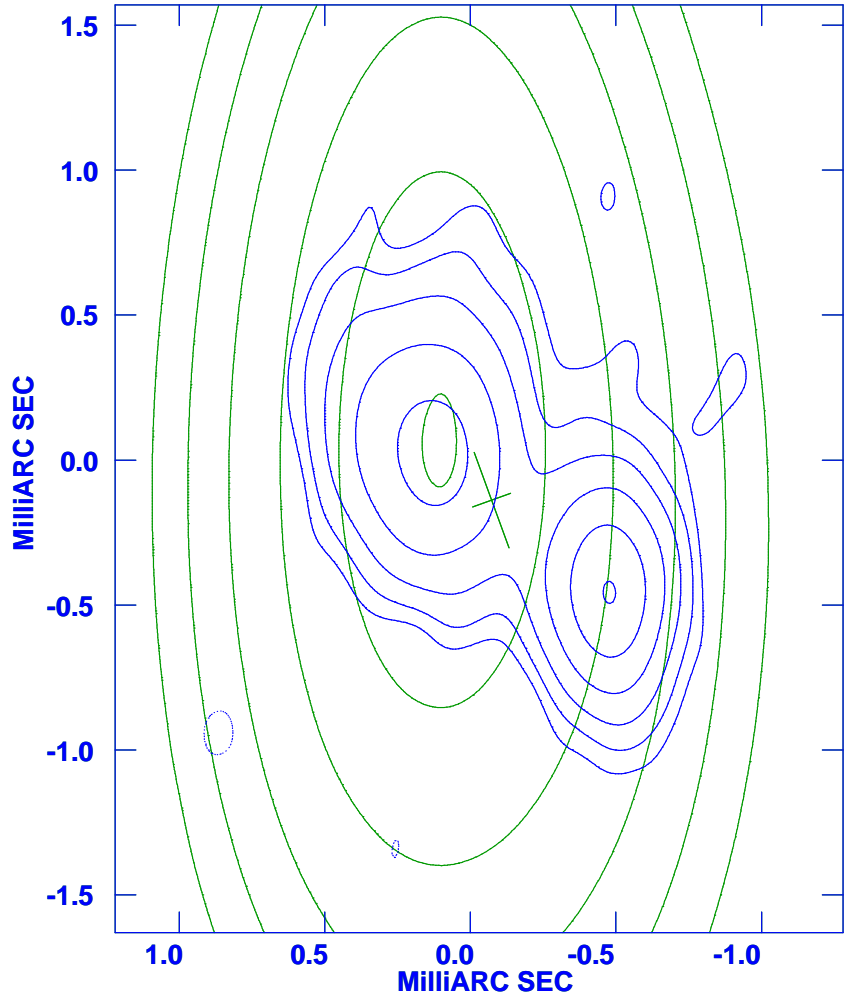}
\vspace{-5cm}
\caption{Frequency-Dependent Structure of 0554+242: The green
  contours show the Jan 2010 8.6-GHz VLBA phase-referencing image.
  The blue contours show the Dec 2009 43 GHz VLBA phase-referencing
  image.  The green error bar gives the estimate of the location of the ICRF2
  position for this source, derived from the registration of the
  0556+238 images.  The contour levels for both images are at 5, 10, 25, 50,
  75, 99\% of the peak.  The tick marks are separated by 0.5 mas.}
\label{fig:0554_registration}
\end{figure}

A comparison of the 43 GHz Dec 2009 and and 8.6 GHz Jan 2010 VLBA
images is given in Fig.~\ref{fig:0554_registration}.  The 8.6 GHz
image, which shows a slight asymmetry to the west, can be modeled with
two Gaussian components that nearly coincide with the 43 GHz component
positions.  The NE component at 8.6 GHz contains 0.73 Jy, compared
with 0.10 and 0.24 Jy at 43 and 23 GHz, respectively.  The SW
component at 8.6 GHz contains 0.13 Jy, compared with 0.07 and 0.13 Jy
at 43 and 23 GHz, respectively.  Hence the steep spectrum of the NE
component and the flat spectrum of the SW component are supported by
the 8.6 GHz model.  Thus, the position (peak brightness) of the radio
source in Jan 2010 is the NE component which is now 0.8 mas north
east of the core.

The ICRF2 position of 0554+242 is located between the two components,
although somewhat closer to the NE component, but is still 0.6 mas
from the core position.  Although no significant motion between the two
components has been detected over the last ten years, the present
ICRF2 position (generated from more than 20 years of data) is
consistent with earlier periods when the NE component may have been
closer to the core, or reflects on the different intensity ratios
between the two components.

\subsection {0601+245 at 8.6 GHz}

\begin{figure}[ht]
\vspace{-6cm}
\hspace{-2cm}
\includegraphics[angle=0,scale=0.9]{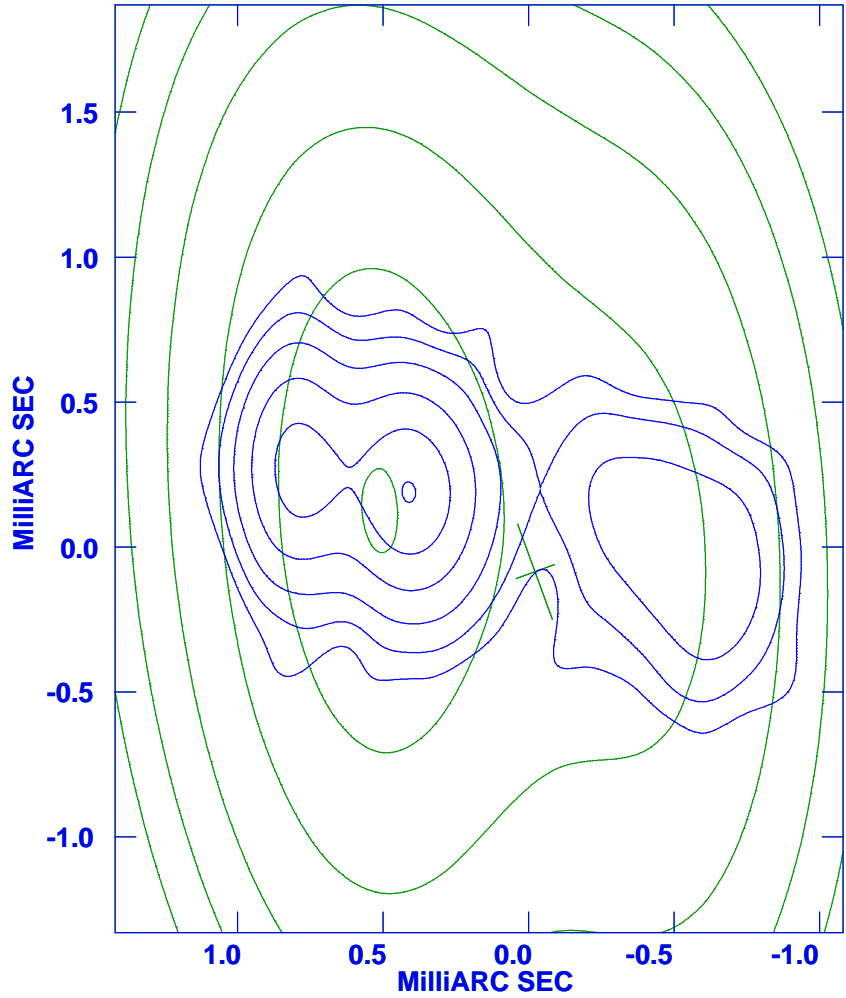}
\vspace{-5cm}
\caption{Frequency-Dependent Structure of 0601+245: The yellow
  contours show the Jan 2010 8.6-GHz VLBA phase-referencing image.
  The blue contours show the Dec 2009 43 GHz VLBA phase-referencing
  image.  The yellow filled ellipse and yellow box (size equal to the
  two-rms error) indicate two possible locations of the ICRF2
  position.  The contour levels for both images are at 5, 10, 25, 50,
  75, 99\% of the peak.  The tick marks are separated by 0.5 mas.}
\label{fig:0601_registration}
\end{figure}

A comparison of the 43 GHz Dec 2009 and and 8.6 GHz Jan 2010 VLBA
images is given in Fig.~\ref{fig:0601_registration}.  The source
structure is complicated, but analysis of the 23 and 43 GHz images
suggests that the radio core is associated with the component at the
eastern part of the emission.  The peak of the 8.6 GHz image (hence
its position) is near the middle component which has a relatively
steep spectral index and would dominate the emission at the lower
frequency.  The western component can be see in the 8.6 GHz image.

The ICRF2 position is displaced significantly to the east, more than
1.0 mas from the radio core, and about 0.6 mas east of the present 8.6
GHz position.  The western component that has a proper motion of about
0.19 mas/yr which extrapolates to a lifetime of about 6 years if the
motion has remained the same.  There are few good quality observations
of this source over the last five years at 8.6 GHz; however, if this
western jet component was stronger in the past, the ICRF2 position
would then have been dominated by this moving component.

\section {Discussion}

The four ICRF2 sources chosen in this study appeared relatively simple
from the 8.6 GHz ICRF monitoring over several decades.  Their emission
was dominated by a compact component and the position error estimates
ranged from 0.05 to 0.15 mas.  The observations at 23 and 43 GHz,
reported here, show that the situation is much more complicated, even
for these high-quality ICRF2 sources.

Although this high resolution, accurate astrometric, multi-frequency,
multi-epoch set of observations covered only four sources, some
general conclusions about the accuracy of the 8.6 GHz ICRF and the
properties of radio cores can be inferred.

\subsection {An Evaluation of the ICRF 8.6 GHz Positions and Structures}

A comparison of the ICRF 8.6 GHz positions and images with those obtained with
the present observations for each of the sources can be summarized as follows:
\begin{description}

\item {\bf 0556+238:} The structure and evolution of this source, one
  of the 295 defining objects of the ICRF2, is not simple even though
  over 90\% of the emission is contained with an angular size of 0.5
  mas.  Many years of observations show virtually no extended emission
  at 8.6 GHz and perhaps a slight extension at 2.3 GHz to the north
  and west (http://rorf.usno.navy.mil/RRFID/).  During the course of
  this experiment, however, a strong jet component moved from 0.1 to
  0.2 mas from the weaker, but optically thick, radio core.  Since the
  source was used as the phase reference calibrator, it produced
  initially what were anomalous properties of the three target
  sources: their radio cores were moving in the sky!.  This type of
  structure variation within a region of 0.3 mas for this source was
  corroborated by analyzing several other experiments over the last
  seven years.  A lesson learned from the results of this source, is
  that structure changes of angular scale 20\% of the resolution can
  occur, and are particularly injurious in determining accurate radio
  positions.

\item {\bf 0547+234:} This source contains about 80\% of its emission
  in a compact component between 8.6 and 43 GHz.  However, the
  position at 8.6 GHz lies about 0.3 mas of the radio core at 23 and
  43 GHz.  The faint emission, observed at all frequencies toward the
  west of the core, may be a diffuse component that becomes
  progressively optically thick at lower frequencies.  Unlike the
  other three sources, this extended emission is not confined to
  well-defined components, but is more ephemeral in time and space.
  This radio core separation between the 8.6 GHz and 43 GHz
  frequencies is similar to that seen for other radio cores
  \citep{kov09}.  However, the other three radio sources contain most
  of their emission in well-confined components and similar core
  shifts associated with an optically thick diffuse emission extending
  from the core do not occur.

\item {\bf 0554+242:} The source structure is relatively simple from
  the 23 GHz and 43 GHz observations, containing a compact, radio core
  to the south-west and a (possibly) slow moving jet component to the
  north-east.  The present 8.6 GHz location coincident with the jet
  component since it is the brighter component at 8.6 GHz.  Thus, the
  present position is displaced currently (Jan 2010) about 0.8 mas
  from the core.  The cataloged ICRF2 position (based on 10 to 15 year
  average position ) is located somewhat closer to the radio core for
  two possible reasons: The NE components was closer to the core 15
  years ago, and/or the flux density of the radio core was more
  dominant.

\item {\bf 0601+245:} The source is relatively large and complex, and
  even the 8.6 GHz-images show an extension to the west.  Only with
  the 43 GHz VLBA observations could the radio core be identified as
  the most eastern component of the three.  The present 8.6 GHz
  position is dominated by the bright, steep-spectrum component that
  is about 0.5 mas west of the core.  The ICRF2 positions is further
  displaced by 0.7 mas to the west.  Unfortunately, there are no
  previous observations for which an accurate structure at 8.6 GHz be
  obtained.  It is suggested that the western component, which is
  moving 0.2 mas to the west, was much brighter ten years ago and this
  component influenced the ICRF2 position much more than presently.

\end{description}

Several conclusions from the study of the four sources are: First, the
position of the source at 8.6 GHz (whether from the VLBA Jan 2010
observation or the ICRF2) for these relatively compact sources is
displaced 0.2 to 1.0 mas from the radio core.  The displacement is
usually in the direction of the more extended structure.  Thus, the
8.6 GHz radio positions, often with an estimated error of $<0.1$ mas,
will often not include the position of the radio core.  The offset may
be caused by the optical depth affect of a relatively smooth jet
connected with the radio core (0547+234 may be an example) and is
usually designated as a core-shift.  However, for the other three
sources, the 8.6 GHz position is influenced by the strong emission
from a discrete component that is not the radio core.

Second, the components that are associated with the 8.6 GHz positions
may move away from the core, and hence the 8.6 GHz positions also
change with time.  This change was noted for two of the sources where
the 8.6 GHz positions measured in Jan 2010 significantly differ from
those of the ICRF2 that are average positions for a source over the
last 10 to 15 years.  Similar changes (small linear motions) in the
position of ICRF sources have been observed for many ICRF2 sources
over the decades of monitoring \citep{FV03}, and the present
observations suggest that such position changes are structure-induced,
rather than intrinsic to the radio source (i.e. radio core) position.

\subsection {Identifying the Radio Cores}

From analysis of 43 and 23 GHz VLBA phase-referenced images of four
source observed over a year, the location of their radio cores have
been ascertained.  The core component for each of the four sources had
some or all of the following properties: the following properties: (1)
The most compact component with 0.3-mas resolution; (2) The component
with the flattest spectral index; (3) The component at one end of the
emission region; (4) The component that is stationary with respect to
the radio cores in other sources.

These criteria could be evaluated by observing the sources in the
following way:

\begin{description}
\item {\bf 43 GHz Resolution of 0.3 mas:} The long VLBA baselines at
  43 GHz were crucial in resolving and determining the radio cores.
  The resolution of 0.3 mas is needed to separate close components or
  to resolve out the inner jet.  For two of the sources the VLBA
  resolution at 23 GHz was not sufficient.  A second reason for
  observing at 43 GHz is that the emission, even from the core region
  of most AGN's, is becoming optically thin \citep{kq-survey1,
    kq-survey2}.  Thus, at 43 GHz little emission is expected to be
  contained in opaque components that produce structure changes with
  frequency.  This is supported by the spectral index of -0.7 for
  several of the radio cores, although a few had flat spectral
  indices.

\item {\bf 23 GHz Observations:} A second observing frequency is
  needed to determine the spectral index of each component, as well as
  producing more accurate images than at 43 GHz (better SNR and less
  tropospheric contamination), albeit with less resolution.  The two
  frequencies should not be separated by more than a factor of two in
  frequency in order to lessen resolution effects in the determination
  of the component spectral index and position.  The spectrum of a
  component is one of the indicators of radio core or non-radio core
  emission.  The cores have a spectral index $>-0.7$ while the other
  components tend to have steeper spectra.

\item{\bf Multiple Session:} Multiple sessions, separated by months to
  a few years, are needed to determine which components have
  significant motion.  This clearly aids in the interpretation of the
  source structure and can be used among several sources in order to
  determine which components in several sources are mutually
  stationary.

\item{\bf Phase Referencing:} The most accurate method to determine
  the relative positions between different sources to about 0.02 mas
  (e/w) and 0.03 mas (n/s) is with the phase referencing.  The gain in
  positional accuracy is because the large tropospheric delay
  variations cancel among closely-space sources.  Unfortunately, phase
  referencing cannot be used for sources more than several degrees
  apart because of the tropospheric refraction that varies linearly
  with source-pair separation.  Group-delay referencing can be done
  over a much larger region of the sky, but the positional accuracy
  over a session is likely to be about 0.1 to 0.2 mas, limited by the
  large-scale tropospheric refraction.  As described above, the
  combination of high resolution and the ability to measure the
  relative position of components that were in different sources
  several degrees apart was crucial in finding which of the components
  in the four sources had separations that were stable to the 0.02 mas
  level.  This was strong evidence that they were stationary in the
  sky since there is little reason for the components in four
  independent sources to be precisely co-moving.

\end{description}

\subsection {Radio Core Stability}

\begin{figure}[ht]
\vspace{-0.6cm}
\hspace{0cm}
\includegraphics[angle=270,scale=0.8]{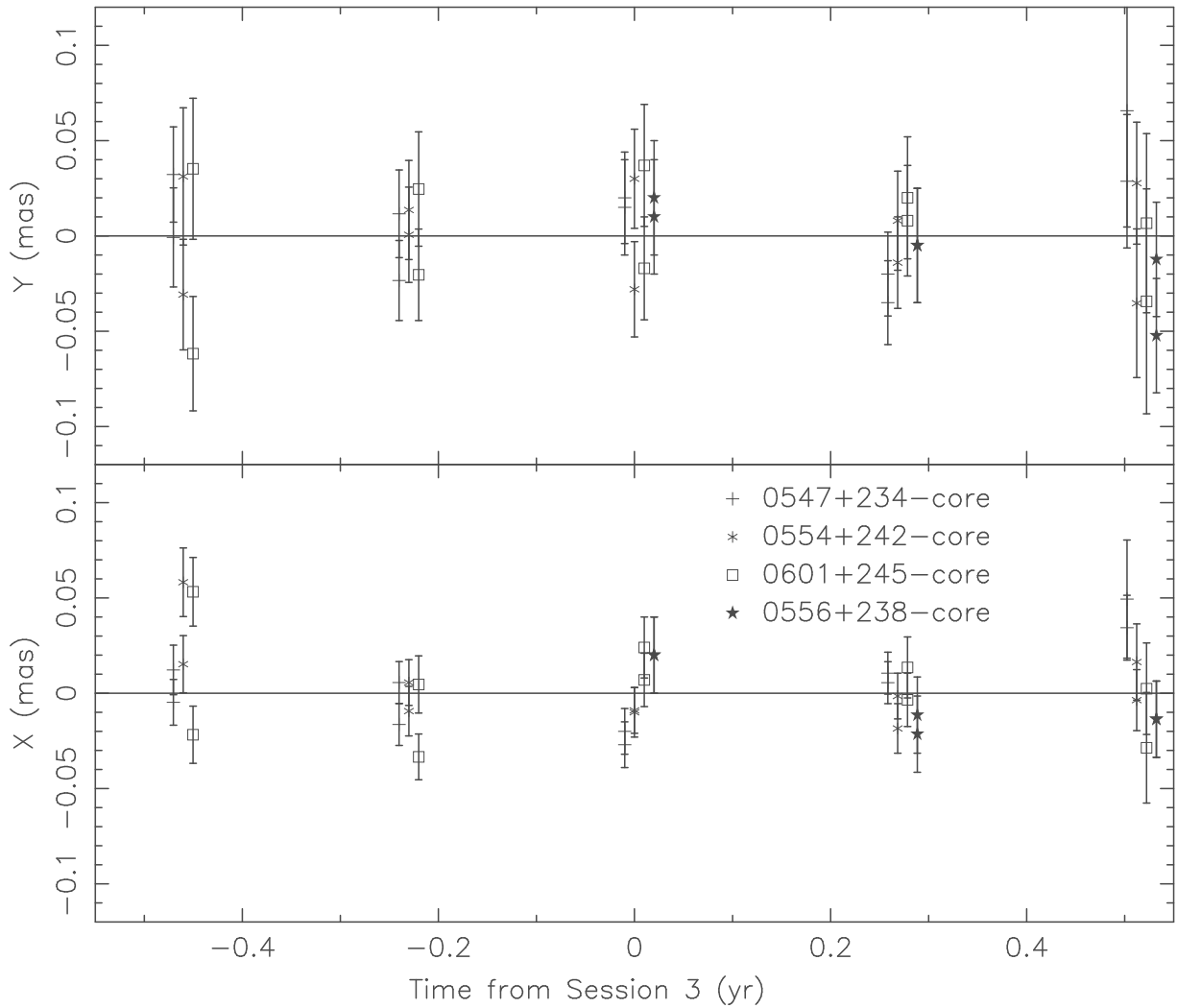}
\vspace{0.5cm}
\caption{The Relative Position of the Radio Cores: The relative
  position of the radio cores of the four sources, after removal of
  their constant position offset, are superposed for the four radio
  cores.  Both 43 GHz and 23 GHz points are plotted separately, and
  the source data is indicated by the symbol.  Each source point is
  slightly displaced in time in order to distinguish among the
  sources.  The third session on Jun 8 2009 was used as the time
  origin.}
\label{fig:resid_motion}
\end{figure}

The measured radio core position stability for the four sources is illustrated
in Fig.~\ref{fig:resid_motion}.  The plot contains the data points from
Fig.~\ref{fig:all_motion}, after removing the proper motion of the
reference grid caused by the motion of the strong jet component in
0556+238.  The data for the SW-component of 0556+238 from
Table~\ref{tab:0556_motion}, when blending with the jet component was
minimal, was also included.  The rms scatter among the sessions and
the sources is dominated by the tropospheric refraction residuals of
about 4 psec (0.02 mas).  The scatter for the first session
(20Dec2008 + 22Dec2008 sessions combined) and the last session is some
larger than the middle three sessions because of lower sensitivity
observations in Dec 2008 and unusually poor weather at many VLBA
locations in Dec 2009.  The normalized $\chi^2$ is 0.7 and 1.6 for the
n/s and e/s offsets, suggesting that the n/s estimated errors are a
factor 0.8 too high and the e/w estimated errors are a factor 1.3 too
low.

There is no significant relative motion for any of the radio cores, and
the rms scatter per point (one radio core for one session) is about
0.02 mas e/w and 0.03 mas n/s.  These sources are separated by an
average of $1.5^\circ$, so a position stability of 0.02 mas of about
one part in $2.7\times 10^8$ was achieved.

Radio cores are believed to be located at the base of the jet which
may be about 2 au from the massive object and accretion disk near the
nucleus of the galaxy.  This separation is about 0.003 mas for the
four sources, assuming a redshift of 1.0.  Whether the radio core
position (and the jet base position) are functions of time is not
known, and one of the goals of these observations are to determine
such changes over time.  At present, the radio cores appear stable to
about 0.02 mas over one year.  Clearly, further observations of these
four sources are needed to determine the longer term stability of
radio cores.  Phase referencing of other sets of sources are also
needed to enlarge the sample of ICRF sources in which the location of
the radio core is known with some assurance.

\subsection {The Source Structure Effect On the ICRF}

The ICRF has determined the position of several hundred sources from
the location of the emission peak associated with compact radio
sources at 8.6 GHz.  The comparison of detailed observations of four
sources at higher resolution and frequency than that of the ICRF
showed that the positions are often displaced up to 0.5 mas from the
radio core, and could vary in position by 0.3 mas/yr.  The ubiquity of
these changes, determined from only four sources, cannot be securely
extrapolated to the defining sources in the ICRF2 catalog as a whole,
but these four sources were among the most compact and brightest of
the ICRF source list.  However, the ICRF2 grid averages the foibles of
each of the 295 sources, so that the average offset or motion to the
inertial grid is decreased by about 15 from any average source
anomaly.  This suggests a stability of 0.02 mas/yr in the orientation
of the grid, and a possible displacement of 0.03 mas in the absolute
position of the grid should be obtained even if the position of many
of the sources are displaced from the radio core in a similar manner
to the four sources studied here.  Hence, the cumulative effect of
source structure induced changes in the ICRF2 grid are comparable to
the errors suggested from the residual tropospheric refraction
residuals that affect the current best modeling of the radio source
positions and other related astrometric/geodetic parameters.

When dealing with individual sources, the estimated errors associated
with the ICRF2 positions are determined from the scatter of many
24-hour session position determinations over several decades.  Some of
the scatter are caused by VLBI network problems and troposphere
modeling errors and tend to be random over periods of months to years.
On the other hand, the apparent position offset of the ICRF2 position
and the radio core, as suggested in this paper, are generally biased
in the direction toward the inner jet which is remains at the same
orientation with the core over many years for most radio sources.
Thus, the estimated ICRF position errors reflect the more random
changes and can be as small as 0.04 mas for sources that have many
observations.  Thus, the offset between the apparent 8.6 GHz position
and the radio core has a systemic component so the offset can be
significantly larger than the estimate position error. \citep{FV03}.

In order to resolve the radio core and to understand the evolution of
an individual radio source, higher frequency observations are needed
to avoid the regions of high optical depth at 8.6 GHz which hide the
location of the true core, and sufficiently high resolution to
resolve and recognize the radio core from other components.  Hence,
imaging the sources in the ICRF2 defining list at 43 GHz (even without
phase referencing) is recommended in order to determine the internal
structure and changes that are often hidden with the 8.6 GHz
observations because of limited resolution and optical depth effects.
These observations would also aid in weeded out some of the 295 defining
sources which have the largest position instability.

\subsection {An ICRF at 23 or 43 GHz}

Phase referencing techniques can determine the accurate structure and
relative position of sources, but only over a small angular region of
several degrees.  Hence, its extension to the entire sky would take a
long period of time and the coupling of phase referencing results over
a few degrees into a whole sky reference frame would probably add
zonal systematic errors that would be hard to remove.

Recently, observations to determine an all-sky catalog at 23 and 43
GHz were initiated.  Part of the motivation was to establish a catalog
of source positions for future spacecraft tracking (at 33 GHz), and
part was to determine if improved accuracy could be obtained at 33 GHz
compared with 8.6 GHz.  The results are given in \citep{kq-survey1,
  kq-survey2}, with most of the observations were made at 23 GHz.  The
basic observable was the group delay, and similar reduction and
analysis techniques, as that for the ICRF, were used.

Although images were made for each source for the twelve observing
sessions, they were not analyzed as completely as those in this paper.
Many sources were dominated by a strong compact component with the
source position given as the peak location of this component.
However, some of these components could be resolved into
sub-components, one of which might be the radio core that could be
displaced more than 0.1 mas from the cataloged position.
Nevertheless, these results strongly suggest that observations at 23
and/or 43 GHz will likely produce images for which a radio core can be
identified.  The analysis of the group delay can then be modified to
remove the structure effects and determine the position for the radio
core component and not simply the position of the maximum emission.
However, until the tropospheric modeling can be significantly
improved, it will be the main limitation to positions derived at 32
GHz for any single session, and with somewhat minimal gains even with
incorporating the imaging information.

\subsection{VLBI2010}

The project VLBI2010 \citep{nie10} is the next generation radio array
that will be used primarily to determine the celestial and terrestrial
reference frames significantly more accurately than the present frame.
Two improvements in observing technique will be used: First, the
telescopes in the array will be small so that they can slew over large
areas of sky very quickly.  This will permit a more accurate
determination of the tropospheric refraction model over the sky more
quickly and with more angular dependence.  Since the troposphere is
the major limitation to the present ICRF, this alone will produce a
significant gain in accuracy, perhaps by a factor of two.

The second improvement will be to observe in a continuous band from
2.5 to 15 GHz and with sufficient SNR in order to convert the measured
group delays into unambiguous phase delays.  Because of the
significant ionospheric refraction component and the changing
structure and resolution of the sources with frequency, the fitting of
the phase delays from these multi-frequency observations will be
complicated.  In other words, the basic observable, the delay (linear
phase slope with frequency) will be contaminated by the ionosphere
delay (phase versus inverse frequency) and the frequency-dependent
source structure phase.  This phase change will consist of the
apparent shifting of the source position with frequency as well as
changes of its internal structure.  Because the highest planned
observing frequency is 15 GHz, it is likely that there will not
sufficient resolution to resolve the radio core within the observed
structure.  The multi-frequency images may help to determine which
part of the source has the flattest spectral index which could
possible identify the radio core.

It is recommended, as it is with the current 8.6 GHz ICRF, that
occasional observations with the VLBA be made at 43 GHz in order to
assess the radio structure at the level of 0.2 mas to better determine
the true position of the radio core.  Another choice may be the
VERA+KVN array, with seven 20-m telescopes, that will begin operation
at 43 GHz by 2012.  The use of the EVN is somewhat limited in its
43-GHz capabilities at the present

\subsection {The Future of the ICRF}

The ICRF as defined by the ICRF-Ext2 \citep{ICRFext2}, and improved
with the ICRF2 \citep{icrf2}, is the realization of the International
Reference System using extragalactic sources (quasars) to establish a
quasi-inertial grid at radio wavelengths.  This is the primary frame
to which the positions of all celestial objects are referenced.  The
295 objects which define the ICRF2 system are somewhat sparsely
distributed over the sky.  Since these objects are typically fainter
than 16-mag, the primary optical catalog, Hipparcos, is referenced to
the radio frame through a few bright quasars and secondarily with
fainter radio stars that are accurately tied to the ICRF-grid
\citep{les98}.  The gaia mission will measure the positions of starts
as faint at 20-mag visual magnitude and will detect over 100,000
quasars in order to produce a large quasar reference frame (LQRF)
\citep{and09}.

Gaia is a scanning instrument that will use quasars to determine the
parallax zero point, the frame spin, and the ICRF frame tie.  It is
estimated that the mission accuracy will be better than 0.025 mas for
stars brighter than 16-mag in the visual, but decreasing to 0.2 mas for
20-mag objects.  With the large number of quasars, Gaia's measurement
should define the inertial reference frame axes to 0.0001 to 0.001
mas.  Since the radio core positions in the ICRF catalog may be in
error by 1 mas, the frame defined by Gaia may be superior to that of
the radio.

The number of precise radio positions has increased dramatically with
the use of the Very Large Array
(VLA)\footnote{http://www.vla.nrao.edu/astro/calib/manual/} and the
VLBA \citep{pet06}.  When combined with the ICRF catalog, there are
over three thousand reference points in the sky.  These catalogs gives
a good densification in the northern hemisphere, and is now increasing
in the south \citep{pet09}.  With this radio densification, comparison
of the Gaia inertial frame with the radio frame may detect zonal
errors in either or both frames.

\section{Conclusions}

Using VLBA phase referencing on four close ICRF2 sources at 23 and 43
GHz for five sessions over one year, we have identified the radio
cores and determined the radio structure of each source with 0.3-mas
resolution, found the spectral index of the radio components, and
measured the motion of the radio components in each source.  The
accuracy of the relative positions of the four sources (through their
radio cores) was about 0.02 mas.  The fours cores showed no relative
motion over the year within the astrometric accuracy.  The sources
contained several other components, some of which were moving up to
1.2 mas/yr from the radio core.

With one VLBA 8.6 GHz session and the ICRF2 positions for these radio
sources, the ICRF2 position determination and stability were assessed.
We found that the ICRF2 position can be dominated by a jet component,
rather than the radio cores that were detected at 43 GHz.  Thus, the
ICRF2 position may be offset from the radio core by up to 0.5 mas.
Since many of the jet components are moving, the ICRF2 positions will
also change, perhaps by 0.2 mas/yr, until the jet component becomes
less intense than the radio core or becomes sufficiently separated
(typically by 1.0 mas).  Such motions have been seen in the ICRF data
base, and all can be attributed to source structure.

Until significantly better tropospheric modeling of ICRF-type
observations can be made, it will be the dominant source of error,
even with these astrometric errors noted in this paper.  Recent VLBA
23 GHz ICRF-type observations have obtained similar accuracies to the
8.6 GHz ICRF2, but more accurate imaging and analysis are needed to
find the radio cores within the brightest components.  The VLBI2010
project shows promise in increasing the accuracy of the reference
frame and other astrometric parameters, but should consider extending
their highest frequency of 15 GHz to 23 or 43 GHz in order to
determine the stable radio core location with more confidence.

\clearpage

\clearpage

\clearpage

\clearpage

\clearpage

\clearpage

\clearpage

\clearpage

\clearpage

\clearpage

\clearpage

\clearpage

\clearpage

\clearpage

\begin{deluxetable}{rrrrrr}
\tabletypesize{\normalsize}
\tablecaption{Source Parameters\label{tab:source_parameters}}
\tablewidth{0pt}
\tablehead{
\colhead {} & \colhead{} & \colhead {} & \multicolumn{3}{c}{Correlated Flux Density} \\
\colhead{Name} & \colhead{RA} & \colhead{DEC} & \colhead{8.6 GHz} & \colhead{23 GHz} &
\colhead{43 GHz} \\
\colhead {} & \multicolumn{2}{c}{J2000} & \colhead {S$^1$~~~~L$^1$} & \colhead {S~~~~~L} & \colhead {S~~~~~L} \\
}
\startdata
0547+234  & 05:50:47.39090  & 23:26:48.1769  &  0.28, 0.25 & 0.24, 0.21 & 0.18, 0.15 \\  
0554+242  & 05:57:04.71358  & 24:13:55.2986  &  0.74, 0.24 & 0.49, 0.27 & 0.22, 0.09 \\ 
0556+238  & 05:59:32.03313  & 23:53:53.9267  &  0.56, 0.42 & 0.32, 0.30 & 0.19, 0.17 \\
0601+245  & 06:04:55.12138  & 24:29:55.0364  &  0.80, 0.34 & 0.43, 0.21 & 0.10, 0.06 \\
\enddata
\vskip -1.0cm
\tablecomments{$^1$: S = 200 km baseline; L = 5000 km baseline}
\end{deluxetable}

\begin{deluxetable}{rrcc}
\tabletypesize{\normalsize}
\tablecaption{Observing Sessions\label{tab:sessions}}
\tablewidth{0pt}
\tablehead{
\colhead {Array} & \colhead{Obs. Date} & \colhead {Freq (GHz)} & 
\colhead{Resol. e/w$\times$n/s (mas)} \\
}
\startdata
VERA  &  18-Apr-2008  &  23      & $1.0\times 1.0$ \\
VERA  &  24-May-2008  &  23      & $1.0\times 1.0$ \\
VLBA  &  20-Dec-2008  &  23,43   & $0.5\times 1.0,~0.3\times 0.6$ \\
VLBA  &  22-Dec-2008  &  23,43   & $0.5\times 1.0,~0.3\times 0.6$ \\
VERA  &  17-Jan-2009  &  23      & $1.0\times 1.0$ \\
VLBA  &  16-Mar-2009  &  23,43   & $0.5\times 1.0,~0.3\times 0.6$ \\
VERA  &  20-Apr-2009  &  23      & $1.0\times 1.0$ \\
VLBA  &  08-Jun-2009  &  23,43   & $0.5\times 1.0,~0.3\times 0.6$ \\
VLBA  &  14-Sep-2009  &  23,43   & $0.5\times 1.0,~0.3\times 0.6$ \\
VLBA  &  12-Dec-2009  &  23,43   & $0.5\times 1.0,~0.3\times 0.6$ \\
VLBA  &  31-Jan-2010  &  8.6     & $1.5\times 3.0$ \\
\enddata
\end{deluxetable}

\begin{deluxetable}{cccrr|ccrr}
\tabletypesize{\scriptsize}
\tablecaption{Structure Parameters for 0556+238, Component {\bf 0}\label{tab:0556}}
\tablewidth{0pt}
\tablehead{
\colhead {\bf Session} & 
\multicolumn{4}{c}{\bf 43 GHz} &
\multicolumn {4}{c}{\bf 23 GHz} \\
& \multicolumn{2}{c}{Intensity [Jy]} &
\multicolumn{2}{c}{Position [mas]} &
\multicolumn{2}{c}{Intensity [Jy]} &
\multicolumn{2}{c}{Position [mas]} \\
& \colhead{Total} & \colhead{Peak} &  
\colhead {X} & \colhead{Y} &
\colhead{Total} & \colhead{Peak}  &  
\colhead {X} & \colhead{Y} \\

}
\startdata
21-Dec-2008 &  0.143 & $ 0.126\pm  0.019$ & $ 0.00\pm 0.01$ & $-0.00\pm 0.02 $ &  0.291 & $ 0.278\pm  0.009$ & $ 0.00\pm 0.01$ & $-0.00\pm 0.02$ \\
16-Mar-2009 &  0.178 & $ 0.162\pm  0.013$ & $-0.00\pm 0.01$ & $ 0.00\pm 0.02 $ &  0.324 & $ 0.311\pm  0.008$ & $ 0.00\pm 0.01$ & $ 0.00\pm 0.02$ \\
08-Jun-2009 &  0.192 & $ 0.175\pm  0.015$ & $-0.00\pm 0.01$ & $-0.01\pm 0.02 $ &  0.320 & $ 0.311\pm  0.009$ & $ 0.00\pm 0.01$ & $ 0.00\pm 0.02$ \\
14-Sep-2009 &  0.205 & $ 0.186\pm  0.014$ & $-0.01\pm 0.01$ & $-0.01\pm 0.02 $ &  0.352 & $ 0.340\pm  0.010$ & $-0.00\pm 0.01$ & $-0.00\pm 0.02$ \\
12-Dec-2009 &  0.217 & $ 0.184\pm  0.017$ & $-0.01\pm 0.01$ & $-0.02\pm 0.02 $ &  0.342 & $ 0.330\pm  0.012$ & $-0.00\pm 0.01$ & $-0.01\pm 0.02$ \\

\hline
\enddata
\end{deluxetable}

\begin{deluxetable}{cccrr|ccrr}
\tabletypesize{\scriptsize}
\tablecaption{Structure Parameters for 0547+234, Component {\bf 0}\label{tab:0547}}
\tablewidth{0pt}
\tablehead{
\colhead {\bf Session} & 
\multicolumn{4}{c}{\bf 43 GHz} &
\multicolumn {4}{c}{\bf 23 GHz} \\
& \multicolumn{2}{c}{Intensity [Jy]} &
\multicolumn{2}{c}{Position [mas]} &
\multicolumn{2}{c}{Intensity [Jy]} &
\multicolumn{2}{c}{Position [mas]} \\
& \colhead{Total} & \colhead{Peak} &  
\colhead {X} & \colhead{Y} &
\colhead{Total} & \colhead{Peak}  &  
\colhead {X} & \colhead{Y} \\

}
\startdata
21-Dec-2008 &  0.187 & $ 0.165\pm  0.045$ & $-0.14\pm 0.01$ & $ 0.24\pm 0.03 $ &  0.330 & $ 0.330\pm  0.049$ & $-0.16\pm 0.01$ & $ 0.27\pm 0.02$ \\
16-Mar-2009 &  0.164 & $ 0.153\pm  0.027$ & $-0.16\pm 0.01$ & $ 0.23\pm 0.02 $ &  0.265 & $ 0.265\pm  0.020$ & $-0.18\pm 0.01$ & $ 0.19\pm 0.02$ \\
08-Jun-2009 &  0.128 & $ 0.115\pm  0.027$ & $-0.19\pm 0.01$ & $ 0.21\pm 0.02 $ &  0.199 & $ 0.199\pm  0.025$ & $-0.20\pm 0.01$ & $ 0.21\pm 0.02$ \\
14-Sep-2009 &  0.162 & $ 0.161\pm  0.026$ & $-0.17\pm 0.01$ & $ 0.15\pm 0.02 $ &  0.227 & $ 0.227\pm  0.020$ & $-0.17\pm 0.01$ & $ 0.13\pm 0.02$ \\
12-Dec-2009 &  0.199 & $ 0.185\pm  0.088$ & $-0.15\pm 0.02$ & $ 0.18\pm 0.03 $ &  0.284 & $ 0.284\pm  0.164$ & $-0.14\pm 0.03$ & $ 0.21\pm 0.06$ \\
\hline
\enddata
\end{deluxetable}

\begin{deluxetable}{cccrr|ccrr}
\tabletypesize{\scriptsize}
\tablecaption{Structure Parameters for 0554+242\label{tab:0554}}
\tablewidth{0pt}
\tablehead{
\colhead {\bf Session} & 
\multicolumn{4}{c}{\bf 43 GHz} &
\multicolumn {4}{c}{\bf 23 GHz} \\
& \multicolumn{2}{c}{Intensity [Jy]} &
\multicolumn{2}{c}{Position [mas]} &
\multicolumn{2}{c}{Intensity [Jy]} &
\multicolumn{2}{c}{Position [mas]} \\
& \colhead{Total} & \colhead{Peak} &  
\colhead {X} & \colhead{Y} &
\colhead{Total} & \colhead{Peak}  &  
\colhead {X} & \colhead{Y} \\

}
\startdata
Component {\bf 0} \\
21-Dec-2008 &  0.103 & $ 0.075\pm  0.037$ & $-0.38\pm 0.02$ & $-0.36\pm 0.04 $ &  0.133 & $ 0.115\pm  0.025$ & $-0.42\pm 0.01$ & $-0.42\pm 0.03$ \\
16-Mar-2009 &  0.087 & $ 0.080\pm  0.023$ & $-0.46\pm 0.01$ & $-0.39\pm 0.03 $ &  0.148 & $ 0.130\pm  0.019$ & $-0.44\pm 0.01$ & $-0.41\pm 0.02$ \\
08-Jun-2009 &  0.097 & $ 0.091\pm  0.026$ & $-0.46\pm 0.01$ & $-0.40\pm 0.03 $ &  0.151 & $ 0.140\pm  0.020$ & $-0.46\pm 0.01$ & $-0.46\pm 0.02$ \\
14-Sep-2009 &  0.089 & $ 0.083\pm  0.023$ & $-0.48\pm 0.01$ & $-0.45\pm 0.03 $ &  0.146 & $ 0.134\pm  0.019$ & $-0.46\pm 0.01$ & $-0.47\pm 0.02$ \\
12-Dec-2009 &  0.067 & $ 0.062\pm  0.026$ & $-0.47\pm 0.02$ & $-0.45\pm 0.03 $ &  0.126 & $ 0.121\pm  0.041$ & $-0.45\pm 0.02$ & $-0.51\pm 0.04$ \\
\hline
Component {\bf 1} \\
21-Dec-2008 &  0.113 & $ 0.058\pm  0.036$ & $ 0.23\pm 0.02$ & $ 0.14\pm 0.04 $ &  0.348 & $ 0.273\pm  0.025$ & $ 0.14\pm 0.01$ & $ 0.08\pm 0.02$ \\
16-Mar-2009 &  0.117 & $ 0.061\pm  0.022$ & $ 0.13\pm 0.01$ & $ 0.08\pm 0.03 $ &  0.333 & $ 0.249\pm  0.018$ & $ 0.15\pm 0.01$ & $ 0.10\pm 0.02$ \\
08-Jun-2009 &  0.124 & $ 0.062\pm  0.025$ & $ 0.15\pm 0.02$ & $ 0.12\pm 0.03 $ &  0.344 & $ 0.261\pm  0.020$ & $ 0.14\pm 0.01$ & $ 0.07\pm 0.02$ \\
14-Sep-2009 &  0.115 & $ 0.063\pm  0.022$ & $ 0.13\pm 0.01$ & $ 0.06\pm 0.03 $ &  0.325 & $ 0.245\pm  0.018$ & $ 0.15\pm 0.01$ & $ 0.06\pm 0.02$ \\
12-Dec-2009 &  0.095 & $ 0.050\pm  0.025$ & $ 0.14\pm 0.02$ & $ 0.05\pm 0.04 $ &  0.284 & $ 0.209\pm  0.040$ & $ 0.16\pm 0.01$ & $ 0.01\pm 0.03$ \\
\hline
\enddata
\end{deluxetable}

\begin{deluxetable}{cccrr|ccrr}
\tabletypesize{\scriptsize}
\tablecaption{Structure Parameters for 0601+245\label{tab:0601}}
\tablewidth{0pt}
\tablehead{
\colhead {\bf Session} & 
\multicolumn{4}{c}{\bf 43 GHz} &
\multicolumn {4}{c}{\bf 23 GHz} \\
& \multicolumn{2}{c}{Intensity [Jy]} &
\multicolumn{2}{c}{Position [mas]} &
\multicolumn{2}{c}{Intensity [Jy]} &
\multicolumn{2}{c}{Position [mas]} \\
& \colhead{Total} & \colhead{Peak} &  
\colhead {X} & \colhead{Y} &
\colhead{Total} & \colhead{Peak}  &  
\colhead {X} & \colhead{Y} \\

}
\startdata
Component {\bf 0} \\
21-Dec-2008 &  0.053 & $ 0.053\pm  0.027$ & $ 0.84\pm 0.02$ & $ 0.35\pm 0.04 $ &  0.139 & $ 0.139\pm  0.031$ & $ 0.76\pm 0.01$ & $ 0.25\pm 0.03$ \\
16-Mar-2009 &  0.065 & $ 0.065\pm  0.024$ & $ 0.78\pm 0.02$ & $ 0.32\pm 0.03 $ &  0.147 & $ 0.147\pm  0.019$ & $ 0.75\pm 0.01$ & $ 0.27\pm 0.02$ \\
08-Jun-2009 &  0.068 & $ 0.068\pm  0.028$ & $ 0.80\pm 0.02$ & $ 0.31\pm 0.03 $ &  0.127 & $ 0.127\pm  0.024$ & $ 0.78\pm 0.01$ & $ 0.25\pm 0.03$ \\
14-Sep-2009 &  0.058 & $ 0.058\pm  0.024$ & $ 0.78\pm 0.02$ & $ 0.27\pm 0.03 $ &  0.114 & $ 0.114\pm  0.024$ & $ 0.76\pm 0.01$ & $ 0.25\pm 0.03$ \\
12-Dec-2009 &  0.039 & $ 0.039\pm  0.033$ & $ 0.76\pm 0.03$ & $ 0.21\pm 0.05 $ &  0.092 & $ 0.092\pm  0.051$ & $ 0.73\pm 0.03$ & $ 0.19\pm 0.06$ \\
\hline
Component {\bf 1} \\
21-Dec-2008 &  0.070 & $ 0.070\pm  0.027$ & $ 0.47\pm 0.02$ & $ 0.24\pm 0.03 $ &  0.190 & $ 0.190\pm  0.031$ & $ 0.41\pm 0.01$ & $ 0.19\pm 0.03$ \\
16-Mar-2009 &  0.068 & $ 0.068\pm  0.024$ & $ 0.42\pm 0.01$ & $ 0.22\pm 0.03 $ &  0.165 & $ 0.165\pm  0.019$ & $ 0.37\pm 0.01$ & $ 0.19\pm 0.02$ \\
08-Jun-2009 &  0.061 & $ 0.061\pm  0.028$ & $ 0.44\pm 0.02$ & $ 0.21\pm 0.03 $ &  0.175 & $ 0.175\pm  0.024$ & $ 0.39\pm 0.01$ & $ 0.17\pm 0.02$ \\
14-Sep-2009 &  0.069 & $ 0.069\pm  0.024$ & $ 0.40\pm 0.01$ & $ 0.19\pm 0.03 $ &  0.167 & $ 0.167\pm  0.024$ & $ 0.37\pm 0.01$ & $ 0.16\pm 0.02$ \\
12-Dec-2009 &  0.052 & $ 0.052\pm  0.033$ & $ 0.39\pm 0.02$ & $ 0.13\pm 0.04 $ &  0.156 & $ 0.156\pm  0.051$ & $ 0.34\pm 0.02$ & $ 0.05\pm 0.04$ \\
\hline
Component {\bf 2} \\
21-Dec-2008 &  0.056 & $ 0.029\pm  0.026$ & $-0.34\pm 0.03$ & $ 0.13\pm 0.06 $ &  0.136 & $ 0.080\pm  0.029$ & $-0.32\pm 0.02$ & $ 0.10\pm 0.04$ \\
16-Mar-2009 &  0.067 & $ 0.030\pm  0.023$ & $-0.44\pm 0.02$ & $ 0.06\pm 0.05 $ &  0.139 & $ 0.104\pm  0.018$ & $-0.44\pm 0.01$ & $ 0.04\pm 0.03$ \\
08-Jun-2009 &  0.064 & $ 0.026\pm  0.026$ & $-0.44\pm 0.03$ & $ 0.06\pm 0.06 $ &  0.125 & $ 0.089\pm  0.023$ & $-0.46\pm 0.02$ & $ 0.03\pm 0.03$ \\
14-Sep-2009 &  0.060 & $ 0.022\pm  0.022$ & $-0.51\pm 0.03$ & $-0.01\pm 0.06 $ &  0.117 & $ 0.077\pm  0.023$ & $-0.50\pm 0.02$ & $ 0.02\pm 0.04$ \\
12-Dec-2009 &  0.047 & $ 0.023\pm  0.033$ & $-0.59\pm 0.04$ & $-0.06\pm 0.09 $ &  0.114 & $ 0.061\pm  0.049$ & $-0.57\pm 0.04$ & $-0.05\pm 0.08$ \\
\hline
\enddata
\end{deluxetable}

\begin{deluxetable}{cccrr|ccrr}
\tabletypesize{\scriptsize}
\tablecaption{Double Structure Parameters for 0556+238\label{tab:0556_motion}}
\tablewidth{0pt}
\tablehead{
\colhead {\bf Session} & 
\multicolumn{4}{c}{\bf 43 GHz} &
\multicolumn {4}{c}{\bf 23 GHz} \\
& \multicolumn{2}{c}{Intensity [Jy]} &
\multicolumn{2}{c}{Position [mas]} &
\multicolumn{2}{c}{Intensity [Jy]} &
\multicolumn{2}{c}{Position [mas]} \\
& \colhead{Total} & \colhead{Peak} &  
\colhead {X} & \colhead{Y} &
\colhead{Total} & \colhead{Peak}  &  
\colhead {X} & \colhead{Y} \\
}
\startdata
{\bf One Component} \\
21-Dec-2008 &  0.160 & $ 0.125\pm  0.010$ & $ 0.00\pm 0.02$ & $ 0.00\pm 0.03 $ &  0.276 & $ 0.255\pm  0.040$ & $ 0.00\pm 0.02$ & $-0.01\pm 0.03$ \\
16-Mar-2009 &  0.208 & $ 0.163\pm  0.020$ & $ 0.00\pm 0.02$ & $ 0.00\pm 0.03 $ &  0.306 & $ 0.280\pm  0.030$ & $ 0.00\pm 0.02$ & $-0.02\pm 0.03$ \\
\hline
{\bf NE Component} \\
08-Jun-2009 &  0.160 & $ 0.160\pm  0.020$ & $ 0.01\pm 0.02$ & $ 0.03\pm 0.03 $ &  0.280 & $ 0.280\pm  0.040$ & $ 0.01\pm 0.02$ & $ 0.02\pm 0.03$ \\
14-Sep-2009 &  0.180 & $ 0.180\pm  0.020$ & $ 0.01\pm 0.02$ & $ 0.02\pm 0.03 $ &  0.270 & $ 0.270\pm  0.030$ & $ 0.01\pm 0.02$ & $ 0.01\pm 0.03$ \\
12-Dec-2009 &  0.189 & $ 0.189\pm  0.010$ & $ 0.00\pm 0.01$ & $ 0.01\pm 0.02 $ &  0.320 & $ 0.320\pm  0.030$ & $ 0.01\pm 0.02$ & $ 0.02\pm 0.03$ \\
\hline
{\bf SW Component} \\
08-Jun-2009 &  0.070 & $ 0.070\pm  0.010$ & $-0.04\pm 0.02$ & $-0.11\pm 0.04 $ &  0.045 & $ 0.045\pm  0.015$ & $-0.04\pm 0.02$ & $-0.12\pm 0.03$ \\
14-Sep-2009 &  0.048 & $ 0.048\pm  0.007$ & $-0.09\pm 0.02$ & $-0.16\pm 0.03 $ &  0.060 & $ 0.060\pm  0.015$ & $-0.08\pm 0.02$ & $-0.16\pm 0.03$ \\
12-Dec-2009 &  0.070 & $ 0.070\pm  0.010$ & $-0.09\pm 0.02$ & $-0.23\pm 0.03 $ &  0.060 & $ 0.060\pm  0.015$ & $-0.09\pm 0.02$ & $-0.19\pm 0.03$ \\
\hline
\enddata
\end{deluxetable}

\end{document}